\documentstyle[ltwol,psfig]{article}

\def\be{\begin{equation}}
\def\ee{\end{equation}}
\def\bea{\begin{eqnarray}}
\def\eea{\end{eqnarray}}
\def\dmsq{\Delta m^{2}~}

\def\nue{\nu_e}

\def\sinsqtheta{\sin^2 2 \theta~}

\def\gtwid{\mathrel{\raise.3ex\hbox{$>$\kern-.75em\lower1ex\hbox{$\sim$}}}}
\def\ltwid{\mathrel{\raise.3ex\hbox{$<$\kern-.75em\lower1ex\hbox{$\sim$}}}}

\newcommand{\text}[1]{\mbox{#1}}

\begin{document}

\title{Recent Results on Neutrino Oscillations}

\author{J. M. CONRAD}

\address{Nevis Laboratories, P.O. Box 137, 
Irvington, NY 10533, USA\\E-mail: conrad@fnal.gov}


%\draft

\twocolumn[\maketitle\abstracts{This article reviews the many new results from neutrino oscillation searches
which were presented at ICHEP '98.  Exciting indications of
neutrino oscillations have been seen in the solar neutrino deficit, atmospheric
neutrino deficit and LSND excess.   These indications and reported limits on
oscillations are considered.  Attempts to develop a theory which
addresses all of the neutrino oscillation data are discussed.   Some of
the remaining questions and the future experiments which will help answer
these questions are described.}]

This article reviews recent data on neutrino oscillations, 
focussing on the new results presented in Parallel Session 2 of the
International Conference on High Energy Phyics (ICHEP) '98.
Section 1 provides an introduction to the formalism of neutrino
oscillations and practicalities of understanding oscillation results.
Section 2 reviews the present
evidence for neutrino oscillations:
the solar neutrino deficit, atmospheric
neutrino deficit and LSND excess.   Several recent experiments 
have completed extensive searches for oscillations and have
established limits on oscillation parameters based on their
null results.  These are reviewed in section 3. Combining the positive
indications and
the limits permits construction of models for oscillations.  Two
types of models are considered in section 4.  After considering
all of the results, it is clear that the present data raise many
questions.
Some of the future experiments which may help resolve these
questions are described in section 5.

\section{Introduction to Neutrino Oscillations}

The existence of neutrino oscillations would 
require a significant departure from the
Standard Model.    Oscillations imply that lepton flavor number is
not conserved.   Furthermore, at least one neutrino must be
massive, which requires a right-handed partner to the 
neutrino.    
There are various ways to accomodate this extension to the Standard
Model.  For example, one can introduce
iso\-singlet partners (``sterile neutrinos'').

At this point, if neutrinos are massive, we know that their mass is tiny.
Kinematic distributions observed in weak decay can be used to place
limits on the mass of outgoing neutrinos. This method has been used to
obtain upper limits on the $e$, $\mu $, and $\tau $ neutrinos from the decay
channels listed in Tab.~\ref{Direct}. It will be extremely difficult 
experimentally to measure 
$m_{\nu _\mu }$ and $m_{\nu _\tau }$ if these masses are at the 
level of, or smaller than, 
a few eV's.  Further progress in the near future will have to
come through searching for neutrino oscillations.

\subsection{Neutrino Oscillation Formalism}

If neutrinos have mass, it is likely that the mass eigenstates 
are different from the weak interaction eigenstates.
In this case, the weak eigenstates
can be written as mixtures of the mass eigenstates, for example:

\[
\begin{array}{l}
\nu _e=\cos \theta \;\nu _1+\sin \theta \;\nu _2 \\ 
\nu _\mu =-\sin \theta \;\nu _1+\cos \theta \;\nu _2
\end{array}
\]
where $\theta$ is the ``mixing angle.''
In this case, 
a pure flavor (weak) eigenstate born through a weak decay will
oscillate into another flavor as the state propagates in space. This
oscillation is due to the fact that each of the mass eigenstate components
propagates with different frequencies if the masses are different, $\Delta
m^2=\left| m_2^2-m_1^2\right|>0$.  In such a two-component model, 
the oscillation probability for 
$\nu_\mu \rightarrow \nu_e$ oscillations is then given by:
\begin{equation}
\text{Prob}\left( \nu _\mu \rightarrow \nu _e\right) = \sin ^22\theta \;\sin
^2\left( \frac{1.27\;\Delta m^2\left( \text{eV}^2\right) \,L\left( \text{km}%
\right) }{E \left( \text{GeV}\right) }\right) 
  \label{prob}
\end{equation}
where
$L$ is the distance from the source, and $E$ is the neutrino energy.

Most neutrino oscillation analyses consider only two-generation mixing 
scenarios, but
the more general case includes oscillations between all three neutrino
species.   This can be expressed as:
\[
\left( 
\begin{array}{l}
\nu _e \\ 
\nu _\mu \\ 
\nu _\tau
\end{array}
\right) =\left( 
\begin{array}{lll}
U_{e1} & U_{e2} & U_{e3} \\ 
U_{\mu 1} & U_{\mu 2} & U_{\mu 3} \\ 
U_{\tau 1} & U_{\tau 2} & U_{\tau 3}
\end{array}
\right) \left( 
\begin{array}{l}
\nu _1 \\ 
\nu _2 \\ 
\nu _3
\end{array}
\right) 
\]
This formalism is analogous to the quark sector, where strong 
and weak eigenstates are not identical and the resultant mixing is described 
conventionally by a unitary mixing matrix.    The oscillation
probability is then:
\begin{eqnarray}
{\rm Prob}&&\left( \nu _\alpha \rightarrow \nu _\beta \right)=\delta_{\alpha \beta }- \nonumber \\
&& 4\sum\limits_{j>\,i}U_{\alpha \,i}U_{\beta \,i}U^*_{\alpha
\,\,j}U^*_{\beta \,\,j}\sin ^2\left( \frac{1.27\;\Delta m_{i\,j}^2 \,L }{E }%
\right)  \label{3-gen osc}
\end{eqnarray}
where $\Delta m_{i\,j}^2=\left| m_i^2-m_j^2\right| $ .
Note that there are three different 
$\Delta m^2$ (although only two are independent)
and three different mixing angles.

Although in general there will be mixing among all three flavors
of neutrinos, two-generation mixing is often assumed for simplicity.
If the mass scales are quite different ($m_3 >> m_2  >> m_1$ for
example), then the 
oscillation phenomena tend to decouple and the two-generation 
mixing model is a good approximation in limited regions.
In this case, each transition can be described 
by a two-generation mixing equation.
However, it is possible that experimental results
interpreted within the two-generation 
mixing formalism may indicate very different $\dmsq$ 
scales with quite different apparent strengths for the same oscillation.
This is because, as is evident from equation \ref{3-gen osc},
multiple terms involving different mixing strengths and $\Delta m^2$ 
values contribute to the transition probability for $\nu_\alpha \rightarrow
\nu_\beta$.

\begin{table}[tbp] \centering%
\caption{Direct neutrino mass measurements (see Ref. 1) from kinematic 
distributions of weak 
decays.\label{Direct}}
\begin{tabular}{lll}
\hline\hline
{ Neutrino Type} & { Mass Limit} & { Process} \\ \hline
Electron & $<\sim 10$ eV & $^3H\rightarrow {}^3He+e^{-}+\nu _e$ \\ 
Muon & $<170$ keV & $\pi ^{+}\rightarrow \mu ^{+}+\nu _\mu $ \\ 
Tau & $<18.2$ MeV & $\tau \rightarrow 5\pi \left( \pi ^0\right) +\nu _\tau $
\\ \hline\hline
\end{tabular}
\end{table}%

\subsection{Neutrino Oscillation Experiments}

\begin{figure}
\centerline{\psfig{figure=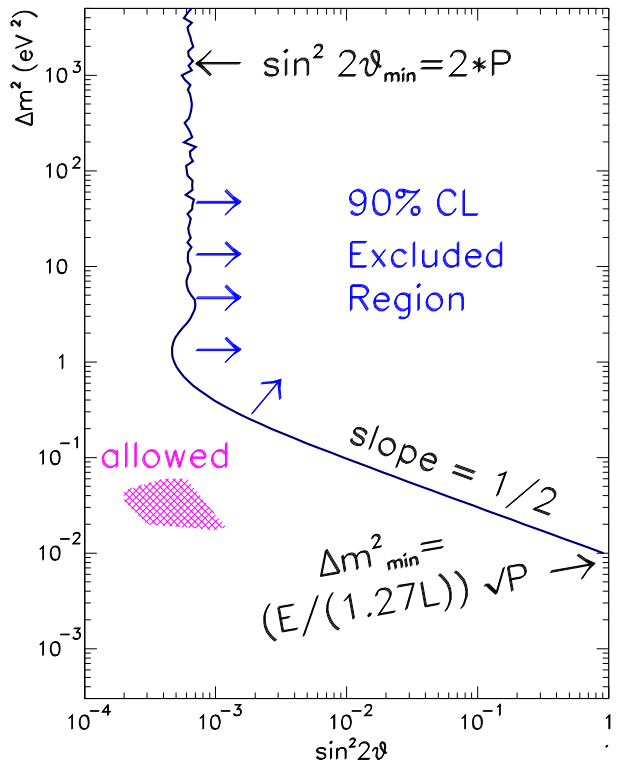,bbllx=100bp,bblly=100bp,bburx=500bp,bbury=700bp,width=3.in,clip=T}}
\caption{Generic example of a neutrino oscillation plot.   The region
to the right of the solid line is excluded at 90\% CL.   The shaded
blob represents an ``allowed'' region.} 
\label{exampleplot}
\end{figure}

From equation \ref{prob}, one can see that 
three important issues confront the designer
of the ideal
neutrino experiment.   First, if one is searching for oscillations 
in the very small $\Delta m^2$ region, then large $L/E$ must be 
chosen in order to enhance the $\sin^2 (1.27 \Delta m^2 L/E)$ term.
However if $L/E$ is too large in comparison to $\Delta m^2$, then  
oscillations occur rapidly.  Because experiments have finite
resolution on $L$ and $E$, and a spread in beam energies, the 
$\sin^2 (1.27 \Delta m^2 L/E)$ averages to $1/2$ when $\Delta m^2 \gg
L/E$ and one loses
sensitivity to $\Delta m^2$.   Finally, because the probability is
directly proportional to $\sin^2 2\theta$, if the mixing angle is
small, then high statistics are required to observe an oscillation
signal.

There are two types of oscillation searches: ``disappearance'' and
``appearance.'' To be simplistic, consider a pure source of neutrinos
of type $x$. In a disappearance experiment, one looks for a deficit in
the expected flux of $\nu_x$. Appearance experiments search for $\nu_x
\rightarrow \nu_y$ by directly observing interactions of neutrinos of
type $y$. The case for oscillations is most persuasive if the deficit
or excess has the ($L/E$) dependence predicted by the neutrino
oscillation formula (equation \ref{prob}).

Let us say that a hypothetical perfect neutrino oscillation experiment sees no
oscillation signal, based on $N$ events.
The experimentors can rule out the probability for oscillations at some
confidence level.   A typical choice of confidence level is 90\%, so
in this case, the limiting probability is $P=1.28 \sqrt{N}/N$.   There
is only one measurement and there are two unknowns, so
this translates to an excluded region within $\Delta m^2$ -- $\sin^2 2\theta$
space.  As shown in Fig.~\ref{exampleplot}, this is indicated by a
solid line, with the excluded region on the right.   At high $\Delta
m^2$, the limit on $\sin^2 2\theta$ is driven by the experimental
statistics.    The $L$ and $E$ of the experiment drive the low
$\Delta m^2$ limit.

The imperfections of a real experiment affect the limits which can 
be set.
Systematic uncertainties in the efficiencies and backgrounds 
reduce the sensitivity of a given experiment.
Background sources introduce multiple flavors of neutrinos 
in the beam.   Misidentification of the interacting neutrino flavor
in the detector can mimic oscillation signatures. In addition, systematic
uncertainties in the relative acceptance versus distance and energy need to
be understood and included in the analysis of the data.  These
systematics are included in the 90\% CL excluded regions presented by
the experiments in this paper.

The most convincing signature for oscillations is a statistically and
systematically significant signal (as opposed to deficit) 
with the dependence on $L$ and $E$ as predicted for oscillations.   
This has not yet been
observed.   Deficits have been observed in the expected rate of two
neutrino sources:  solar and atmospheric.   A signal has been been
observed by the LSND experiment, but it is not at 5$\sigma$
significance and the $L$ and $E$ dependence has not yet been clearly
demonstrated.   

Indications of neutrino oscillations 
appear as allowed regions, indicated by shaded areas (see 
example in Fig.~\ref{exampleplot}), 
on plots of $\Delta m^2$ {\it vs.} $\sin^2 2\theta$.  

\subsection{Small Statistics Experiments with Background}

\begin{figure}
\centerline{\psfig{figure=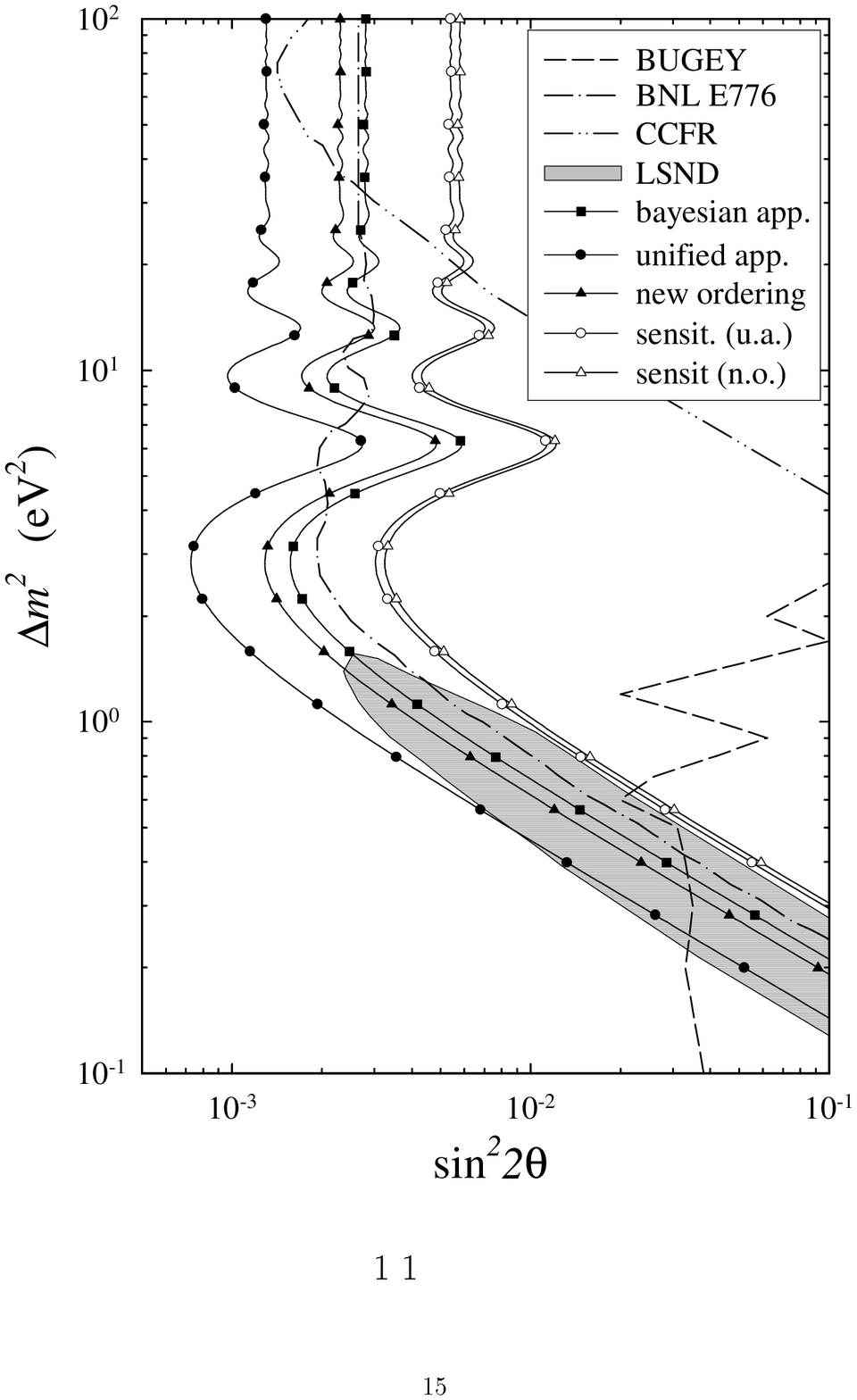,bbllx=100bp,bblly=130bp,bburx=500bp,bbury=700bp,width=2.8in,clip=T}}
\caption{Range of possible interpretations of the KARMEN data.
``Unified app'' refers to the Feldman-Cousins Method, while ``new
ordering'' refers to the Giunti Method.  As per the Feldman-Cousins
recommendation, the sensitivities associated with the ana\-lysis methods
are also shown. (plot from Ref.~4)
} 
\label{giuntips}
\end{figure}

Care must be taken when comparing excluded and allowed regions near
the boundaries. First, one should remember that a 90\% CL exclusion
limit means that if a signal were in this region, 10\% of the time it
would not be seen. Second, there is no consensus within the physics
community on the method for determining allowed regions and 
limits.  Although in most cases, the deviations of the different
methods are small, there are cases for which the disagreements are
significant.  As an example, consider Fig.~\ref{giuntips}, from Ref.~4.  The
details of the experimental results presented in this plot will be
considered later.   At this point, the reader should consider the 
shaded area to be an allowed region from an oscillation
experiment (LSND) and the curves to represent the different statistical 
interpretations of 90\% exclusion regions {\it of the same set of
experimental data} (KARMEN 2).  

In a hypothetical analysis, 
take $\mu$ to be the expected signal and $b$ to be the
average expected background, and $n=\mu+b$ to be the total number of
expected events.    The experiment will observe 
some number of events, $n_{obs}$, which is not necessary equal to $n$.
How is the measurement, $n_{obs}$, related to an estimate of $\mu$, the
true signal? In other words, what can $n_{obs}$ tell us about the
underlying physics?

In principle, the method is straightforward.
We can construct two curves, $n_1(\mu,\alpha)$ and 
$n_2(\mu,\alpha)$ such that the probability for $n$ to be
between $n_1$ and $n_2$ is $\alpha$.  Typically, $\alpha$ is chosen to 
be 90\%.   For small statistics, a Poisson probability
distribution is used.  We then invert the relationship, obtaining 
the functions $\mu_1(n,\alpha)$ and $\mu_2(n,\alpha)$.  Thus, 
$n_{obs}$ implies that the signal, $\mu$, lies somewhere between 
$\mu_1(n_{obs},\alpha)$ and $\mu_2(n_{obs},\alpha)$ 
with probability $\alpha$.
In practice, the construction of  $n_1(\mu,\alpha)$ and $n_2(\mu,\alpha)$
can be done in various ways, and this forms the heart of this controversy.

The method which has been accepted in the past, and was previously endorsed by the 
Particle Data Group,\cite{oldpdg} 
has an inherent 
inconsistency, as pointed out by Feldman and Cousins.\cite{feldman}
In this method, $n_1$ is chosen such that $n<n_1(\mu,\alpha)$ with
probability $(1-\alpha)/2$ and $n_2$ is chosen such that
$n>n_2(\mu,\alpha)$  with probability $(1-\alpha)/2$.   This gives you 
a central confidence interval based on a two sided Gaussian
distribution in the case of a signal.  But in the case of a limit, you
have only a one sided confidence interval with $n<n_1(\mu,\alpha)$ with
probability $(1-\alpha)$.  Thus your treatment of the data ``flip-flops,''
to use the terminology of Feldman-Cousins, depending on whether you
are setting a limit or determining a signal region.   

Feldman-Cousins developed a new method which is now adopted by the
Particle Data Group, called ``The Unified Approach.'' In this
approach, you order the $n$ possibilities by the function
$R(n) = {\rm Prob}(n|\mu b)/{\rm Prob}(n|\mu_{best} b)$ where
$\mu_{best}(n)=\mu$ with maximum probability for $n$ and $b$.  
The technique is to
first order the  $R(n)$'s from largest to smallest (thus the method is based
on ``likelihood ordering''),  then sum the $R(n)$
values until you reach Prob$=\alpha$.    

This Unified Approach has good features and drawbacks.   By design,
there is now a smooth transition between limits and signals -- no more 
flip-flopping.  But, if an experiment sees a number of events smaller
than background, then this method sets a stringent upper bound on
$\mu$, {\it not because of sensitivity to small signals but because of
the fact that too few background events are observed}.

Feldman-Cousins recognized this problem in their paper\cite{feldman} and raised the 
questions:
``Why should an experiment claim credit for expected backgrounds when 
it is clear, in that particular experiment, there were none?  Or why
should a well-designed experiment which has no background and observes
no events be forced to report a higher upper limit than a less
well-designed experiment which expects backgrounds, but, by chance,
observes none?''     The Feldman-Cousins prescription for this
quandary is that an
experiment which sees a low background fluctuation quote both the 
limit and the experimental sensitivity, where ``sensitivity'' is
defined as the average expected limit if an experiment 
with no true signal, only background, were performed
many times.  This lets readers draw 
conclusions based their own personal opinion of what is 
acceptable.

In a recent paper,\cite{giunti} Giunti proposes a new
method of ordering classical confidence intervals which guarantees a
smooth transition from the limit to signal region while resulting in a
weaker improvement in confidence level for the case where the  
background fluctuates low.   In this case, $\mu_{best}(n)$ is
replaced by 
$\mu_{ref}(n)=\int_0^\infty \mu P(\mu|n,b) d\mu$ where 
$ P(\mu|n,b)$ is the Bayesian probability distribution for 
a constant $\mu \ge 0$.  The data are then likelihood ordered
according to $R(n) = {\rm Prob}(n|\mu b)/{\rm Prob}(n|\mu_{ref} b)$.

\begin{figure}
\centerline{\psfig{figure=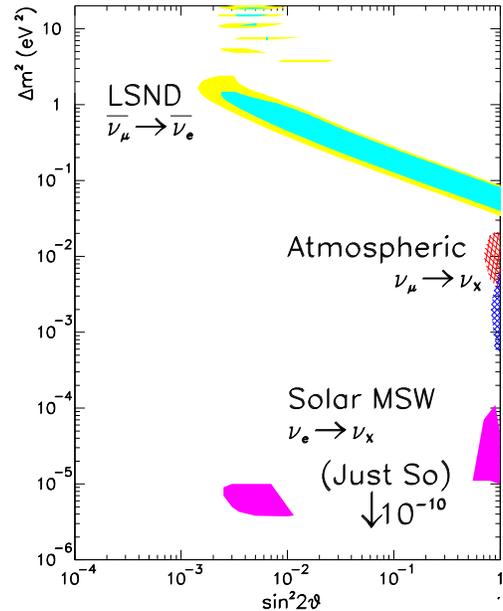,bbllx=0bp,bblly=100bp,bburx=575bp,bbury=600bp,width=3.5in,clip=T}}
\caption{Hints for oscillations come from three sources: solar
neutrinos, atmospheric neutrinos and accelerator produced neutrinos
(LSND).   The solar neutrino deficit has two possible oscillation solutions, MSW
and Just-So (see text).  Allowed regions for these three indications are presented here.
Note that some regions are already addressed by exclusion experiments
discussed in section 3.} 
\label{allhints}
\end{figure}

Comparison of the two methods demonstrates the relative stability of
the limit
with background fluctuations.
Consider 
Fig.~\ref{giuntips}, which addresses an experiment in which $2.88\pm0.013$ 
background events are expected and zero events are observed.
The two recent methods calculate very similar sensitivities of the experiment,
as indicated by the open symbols.   The Unified Approach
(Feldman-Cousins) experimental sensitivity is 4.4 events while the 
New Ordering Approach (Giunti) sensitivity is 4.9 events.
The limits are radically different, however.  The
Unified Approach sets the limit as 
$\mu_{90\% CL}=1.1$~events,  
which is better than
the sensitivity by better than a factor of four.   The New Ordering
Approach sets a limit at 1.9 events.  This is not far from the sensitivity.  

A more conservative alternative is to use a Bayesian
approach.  Arguably, if less background is observed than expected,
then one may be overestimating the background.  Therefore one should 
set a limit assuming that $n_{obs}=b$, in this case, zero.
Therefore this method sets the limit at 2.3 events, as shown on
Fig.~\ref{giuntips}.

In summary, there is disagreement about how to handle data in the case
of background plus a small expected signal.
If an experiment sees the expected background, then the differences in
the statistical methods are relatively small, but when the background
fluctuates low there can be significant differences in limits.
The reader should be skeptical of strong
conclusions drawn in cases of small statistics.     
The goal now should be high statistics, low systematics experiments
designed to address the hints for oscillations which have been observed.

\section{Experiments Reporting Evidence for Oscillations}

Three  different sources of neutrinos have shown
deviations from the expectation, consistent with oscillations. 
The first, called the ``Solar Neutrino Deficit,'' is a low rate of
observed $\nu_e$'s from the sun.   The data are consistent with
$\Delta m^2 \sim 10^{-10} {\rm eV}^2$ or 
$\Delta m^2 \sim 10^{-5} {\rm eV}^2$, depending on the theoretical
interpretation.    The second, called the
``Atmospheric Neutrino Deficit,'' refers to neutrinos produced by 
decays of mesons from cosmic ray interactions in the atmosphere.  
An observed anomalously low ratio
of $\nu_\mu/\nu_e$  can be interpreted as oscillations with $\Delta
m^2 \sim 10^{-3} {\rm eV}^2$.  The third observation is an excess of
$\overline{\nu}_e$ events in a $\overline{\nu}_\mu$  beam by the LSND
experiment, with $\Delta m^2 \sim 10^{-1} eV^2$.  Fig.~\ref{allhints}
summarizes the allowed regions from these results.
In this section each 
case is considered in detail.

\subsection{The Solar Neutrino Deficit}

\begin{figure}
\centerline{\psfig{figure=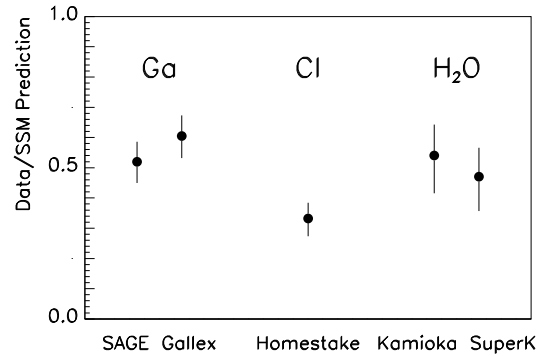,bbllx=0bp,bblly=450bp,bburx=500bp,bbury=800bp,width=3.in,clip=T}}
\caption{Ratio of measurements from five solar neutrino experiments, to the SSM (BP98) prediction.} 
\label{fig:solarexp}
\end{figure}

As shown in Fig.~\ref{fig:solarexp}, three different types of
experiments have observed fewer neutrinos from the sun than 
expected from the Standard
Solar Model (SSM).\cite{BP98}
The first observation of this $\nue$ deficit was made using a Cl target in
the Homestake mine by 
Davis and collaborators.\cite{davis}   Four additional experiments have
confirmed these observations.  The  GALLEX and SAGE
experiments search for electron neutrino interactions
in a Ga target.\cite{gallium}
The Kamiokande and Super Kamiokande (``Super K'') experiments
observe $\nu_e + e \rightarrow \nu_e  + e$ reactions in water.\cite{skamsun}
The gallium and water target experiments indicate a deficit of $\sim
1/2$, while the Cl experiment is lower.  
Results from GALLEX (D. Vignaud) and Super K (M. Vagins, M. Takita)
are reported in these proceedings. 

The solar data have been gathered over an extended 
period of time and many systematic checks on efficiencies and calibrations
have been performed.  For example, the Super K experiment has
installed a LINAC at the detector for {\it in situ} calibration from 5
to 16 MeV.   As a second example, the GALLEX experiment, 
which presented final results
at this conference, doped their detector with $^{71}{\rm As}$,
which decays to the signature $^{71}{\rm Ge}$.  Recovery was measured
to be 99.8$\pm$0.8\%.   This is one of many source tests performed by
the GALLEX group and described at this conference.

\begin{figure}
\centerline{\psfig{figure=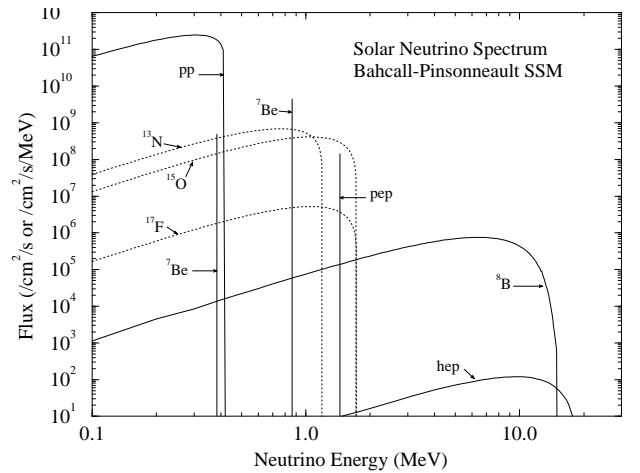,bbllx=0bp,bblly=150bp,bburx=500bp,bbury=550bp,width=3.5in,clip=T}}
\caption{Neutrino fluxes as a function of energy from the sun as
predicted in Ref.~13} 
\label{nuesun}
\end{figure}

\begin{table}[t]

\caption{Fraction of $\nu_e$'s expected from reactions in the Standard
Solar Model for the three types of solar neutrino
experiments.}

\label{solartab}

\vspace{0.4cm}

\begin{center}

\begin{tabular}{|c|c|c|c|}
\hline 
~~& Super K/ & Homestake & GALLEX/SAGE \\
~~& Kamioka &  &  \\
\hline
pp & & & 0.538 \\
$^7$Be I & & & 0.009 \\
$^7$Be II & & 0.150 & 0.264 \\
$^8$Be & 1 & 0.775 & 0.105 \\
pep & & 0.025 & 0.024 \\
$^13$N & & 0.013 & 0.023 \\
$^15$O & & 0.038 & 0.037 \\
\hline
\end{tabular}
\end{center}
\end{table}

Each type of solar experiment has a different energy threshold for
observing $\nu_e$ interactions, and thus is
sensitive to different reactions producing neutrinos in the sun.
The characteristic range of solar $\nu$ energies from each production 
mechanism is shown in Fig.~\ref{nuesun}.
Major sources of solar neutrinos for each experiment are listed in  
Tab.~\ref{solartab}.
The 
``{\it hep}'' ($^3He + p \rightarrow ~^4He + e^+ \nu_e$) process neutrino
contribution is $\sim 10^{-4}$ of the $^8B$ contribution in most solar
models, which is too low to be listed in Tab.~\ref{solartab}, thus 
Super K sees effectively 100\%  $^8B$ neutrinos.   Recently, however,
it has been pointed out that the {\it hep} flux is not well
constrained and could be twenty times larger than 
past models have predicted.\cite{Bahcall_hep}   With this increase,
the {\it hep} neutrinos still remain a small component of the 
Super K flux, but the expected neutrino 
energy distribution changes slightly.  The effect of this shift is
discussed below.

Two important theoretical issues related to the solar neutrino fluxes are the
fusion cross sections and the temperature of the solar interior.
Consensus is developing on 
the systematic uncertainties associated with the dominant 
solar fusion cross sections.
A recent comprehensive analysis\cite{Adelberger} of the available
information on nuclear fusion cross sections important to solar
processes provides the best values along with estimated uncertainties.
These are included in the uncertainty in the SSM value for the ratio
of data to prediction shown in Fig.~\ref{fig:solarexp}.
The flux of neutrinos from certain processes, particularly 
$^8B$, depends strongly
upon temperature.   Recent results in helioseismology have provided an
important test of the SSM.\cite{bahcallseis}   
The sun is a resonant cavity, with
oscillation frequencies dependent upon $U=P/\rho$, the ratio of pressure to
density.   Helioseismological data confirm the SSM prediction of $U$
to better than 0.1\%.\cite{helioseis}  Helioseismological
data are not included in the data used to determine the SSM, therefore
this is an entirely independent cross-check of the model.  The systematic
error associated with temperature dependence is included in the
theoretical error on the SSM prediction.

\begin{figure}
\centerline{\psfig{figure=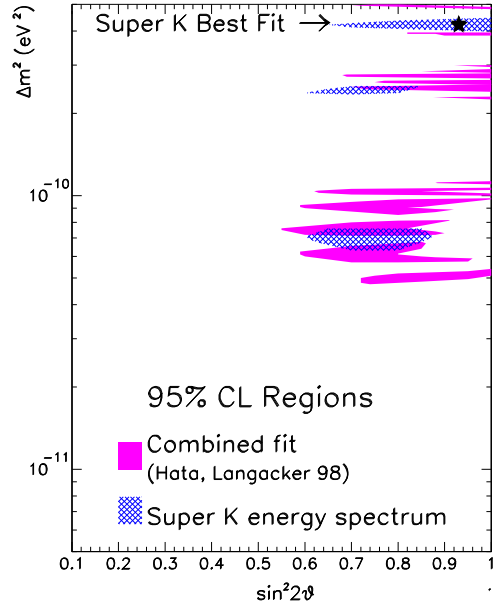,bbllx=50bp,bblly=100bp,bburx=550bp,bbury=670bp,width=3.in,clip=T}}
\caption{Allowed regions in the $\dmsq$ vs $\sinsqtheta$ parameter
space from the four solar neutrino experiments assuming vacuum
oscillations.} 
\label{fig:justso}
\end{figure}

Interpreting the deficit of solar neutrinos as a signal for
oscillations, one calculates the vacuum oscillation probability
using equation \ref{prob}, as shown in Fig.~\ref{fig:justso}.
Vacuum oscillations are often referred to as the ``Just So
Hypothesis'' because this theory assumes that the position of the
earth from the sun is 
at a distance which is an oscillation maximum.
The energy of the neutrinos (only a few MeV) combined with the
long path length from the sun to the earth ($\sim 10^{11}m$) results in allowed
regions of $\Delta m^2$ which are very low ($\Delta m^2 \sim 10^{-10}
{\rm eV}^2$).  In Fig.~\ref{fig:justso},
an analysis of the overall fluxes\cite{hata}
is compared to the recent energy spectrum analysis from Super
K, 
showing that the overlap of these two analyses is limited.
Increasing the {\it hep} neutrino flux, as discussed above,  
modestly improves the agreement between
overall flux and energy spectrum data.\cite{Bahcall_hep}

An alternative oscillation scenario,
referred to as the MSW (Mikheyev-Smirnov-Wolfenstein) solution,\cite{wolf}
includes ``matter effects.''   These effects occur because at low
energies, the electron neutrino has both charged- and 
neutral-current elastic
scattering with electrons, while the $\nu_\mu$ and $\nu_\tau$
experience only neutral-current scattering.
The additional $\nu_e$ interactions
introduce a phase shift as the mass state, which is a combination of
flavor eigenstates,  propagates.   This leads to an
increase in the oscillation probability:
\begin{equation}
\text{Prob}\left( \nu _e \rightarrow \nu _\mu\right) =\left(\sin
^22\theta/W^2 \right) \;
\sin^2\left(1.27 W \Delta m^2 L/E \right) 
  \label{probMSW}
\end{equation}
where 
$W^2 = \sin^2 2\theta + (\sqrt{2}G_F N_e (2E/\Delta m^2) - \cos 2\theta)^2$
and $N_e$ is the electron density. In a vacuum, where $N_e=0$, this
reduces to equation~\ref{prob}.   But within the sun, where the
electron density varies rapidly, ``MSW resonances,'' or large
enhancements of the oscillation probability can occur.    It is also
possible to have matter effects occur as neutrinos travel through the
earth.  For this reason, Super K has searched for a ``day-night
effect'' -- a difference in interaction rate at night due to the MSW
effect in the earth's core.  

\begin{figure}
\centerline{\psfig{figure=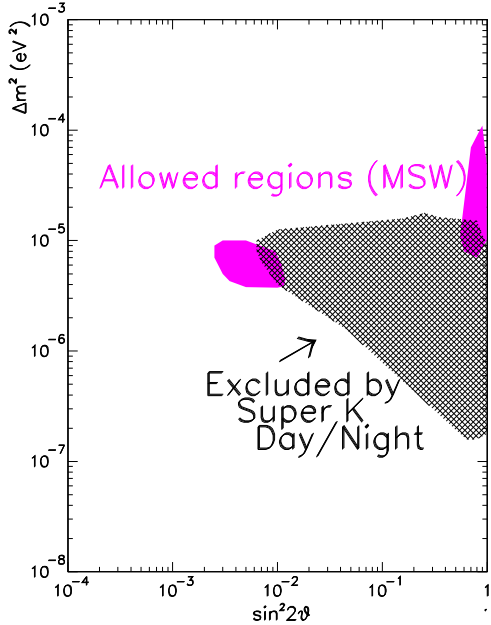,bbllx=50bp,bblly=0bp,bburx=550bp,bbury=700bp,width=3.in,clip=T}}
\caption{Solid: Allowed regions from the solar neutrino experiments, 
including the MSW
effect. Hatched: Excluded (90\% CL) region due to no day-night effect.} 
\label{fig:msw}
\end{figure}

Results of a combined analysis of the solar neutrino data  
within the MSW framework are shown in Fig.~\ref{fig:msw}.  Allowed
regions are indicated by the two solid areas, which are referred to as
the small mixing angle (SMA) solution and large mixing angle (LMA)
solution.   Super K has seen no evidence of a day-night effect.  As a
result, a region indicated by the hatched area can be excluded at 90\% CL.
For more information, see the contribution by M. Vagins in these
proceedings.

The recent data have been compared to the SSM in a global analysis by 
Bahcall, Krastev and Smirnov.\cite{BKS}  The MSW SMA solution 
provided the best fit, with $\Delta m^2 = 5\times 10^{-6} {\rm eV}^2$
and $\sin^2 2\theta = 5.5 \times 10^{-3}$.  The confidence level of
the fit was 7\%.  The confidence level for the LMA solution was
$\sim 1\%$.   The Just-So solution gave a best fit of  
$\Delta m^2 = 6.5 \times 10^{-11} {\rm eV}^2$
and $\sin^2 2\theta = 0.75$, with a 6\% CL.  
Increasing the {\it hep} neutrino flux by a factor of 20 above the SSM
improves the confidence level of all of these fits.\cite{Bahcall_hep}
The SMA solution increases to $\sim$20\%, the LMA solution to 
$\sim$5\% and the ``Just-So'' to $\sim$15\%.  Although the confidence
levels of the fits are relatively low, all three solutions are still 
possible.

The confidence level of fits which do not include oscillations is very
low. The data are inconsistent with the SSM-without-oscillations 
at the 10$\sigma$ level.  
If one allows the flux from each neutrino source to have an arbitrary
normalization, fits can be obtained which are 
inconsistent at only the 3.5$\sigma$ level, but are excluded by the
helioseismology measurements.   Therefore, oscillations appear to 
be the best explanation for the solar neutrino deficit.

\subsection{The Atmospheric Neutrino Deficit}

Neutrinos may be produced in the upper atmosphere.
They result from     
cosmic rays colliding with atmospheric nucleons, 
producing mostly pions which then 
decay into muons and muon-neutrinos.   The resulting muons may also
decay, producing a muon- and an electron-flavored neutrino.  Thus,
this decay chain is expected to produce a two-to-one ratio for
$\nu_\mu$
to $\nu_e$.   

\begin{figure}
\centerline{\psfig{figure=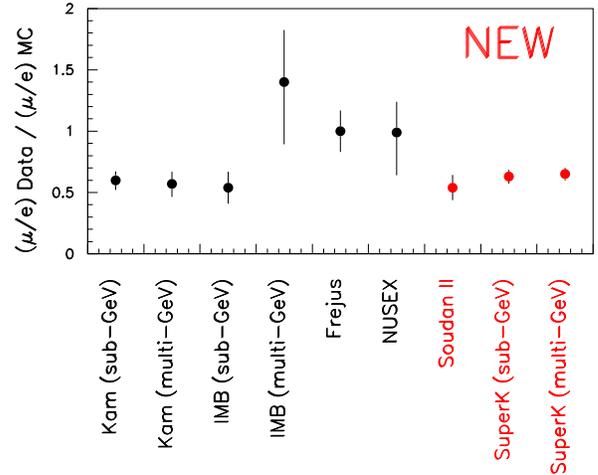,bbllx=50bp,bblly=350bp,bburx=570bp,bbury=800bp,width=3.5in,clip=}}
\caption{$(\nu_\mu/\nu_e)_{data}/(\nu_\mu/\nu_e)_{MC}$ for 
atmospheric experiments.    The three newest results are listed at right.} 
\label{ratofrat}
\end{figure}

Atmospheric neutrinos have been detected through
their charged--current interactions in detectors on the earth's
surface.  
The sensitivity of these detectors to electrons and muons varies over the 
observed energy range.
For example, electron events are mostly contained in the detector, while muon 
events have longer range and escape the detector at higher energies.
Therefore, the results are often divided into 
sub-GeV (contained) and multi-GeV (partially-contained) samples.

The observed ratio of muon to electron neutrino events 
divided by the ratio of events calculated in a Monte Carlo simulation
($R=(\nu_\mu/\nu_e)_{data}/(\nu_\mu/\nu_e)_{MC}$)
for nine atmospheric neutrino analyses is reported in
Fig.~\ref{ratofrat}.  
The newest results 
are  reported in these proceedings by H. Gallagher for Soudan and 
C. McGrew and M. Takita for Super K.  
Within statistics, all experiments are consistent with $R\sim 60\%$.
This deviation from the expected ratio of one is called the 
``Atmospheric Neutrino Deficit.''

\begin{figure}
{\psfig{figure=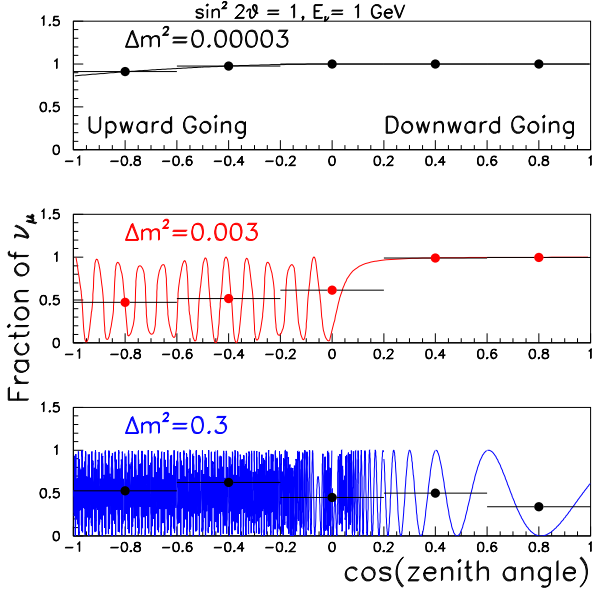,bbllx=0bp,bblly=100bp,bburx=575bp,bbury=700bp,width=3.5in,clip=T}}
\caption{Disappearance of $\nu_\mu$ due to oscillations as a function
of cosine of zenith angle for three regions of $\Delta m^2$.
Resolution of the experiments and limitations of statistics smooth
rapid oscillations such that the data will tend to look like the points} 
\label{costhetz}
\end{figure}

If the atmospheric neutrino deficit is due to neutrino oscillations,
then one would expect a change in $R$ as a function of neutrino path length.
Neutrinos which are produced in interactions directly above the
detector, called ``downward-going,'' traverse
$L\sim10$km.   Neutrinos which are produced on the opposite side of
the earth, called ``upward-going,'' travel
$L\sim 10,000$km.   Traditionally, the path of the neutrino is
described by $\cos \theta_z$, where $\theta_z$ is the zenith angle
measured from directly above the detector.  It should be noted that 
the actual path length, $L$,  is rapidly changing with $\cos \theta_z$.
As an example, the $R$ dependence $vs.$ $\cos \theta_z$ for
oscillations of 1 GeV $\nu_\mu$'s 
which are ``disappearing''  is shown in 
Fig.~\ref{costhetz} for low,
medium and high $\Delta m^2$ values.   At very low $\Delta m^2$, the
probability for oscillation is low, even for very long path lengths, so
$R$ will be consistent with one.  At very high $\Delta m^2$, even the 
neutrinos from above have had the opportunity to oscillate. 
The actual measurement will not resolve the rapid oscillations because
of finite bin sizes, 
resolution of the detector, interaction physics and because the
neutrinos produced in collisions in the atmosphere are not
mono-energetic.  The points on Fig.~\ref{costhetz} are meant to
indicate what might be measured for the three $\Delta m^2$ cases.
A typical experiment would be unable to resolve rapid oscillations.
This makes interpretation of the data, particularly in the moderate
$\Delta m^2$ region, difficult.

\begin{figure}
\centerline{\psfig{figure=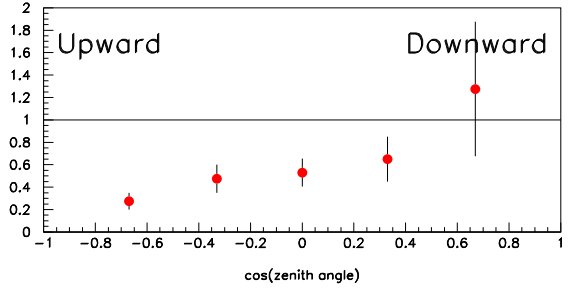,bbllx=0bp,bblly=300bp,bburx=575bp,bbury=700bp,width=3.5in,clip=T}}
\caption{Zenith angle distribution of $R$ from the Kamiokande experiment.}
\label{fig:k2}
\end{figure}

\begin{table*}[t]

\caption{ $\Delta m^2$ from $\cos \theta_z$ dependence of 
atmospheric neutrino experiments using
the 
$\nu_\mu \rightarrow \nu_\tau$ oscillation hypothesis.  
Results that do not have a ``best fit'' quoted by the experiment
are listed with the $\Delta m^2$ which is described as
``consistent with'' the data. Under ``references,'' listed names refer
to contributions to these proceedings.  All data but Kamiokande are 
preliminary.}

\label{tab:deltam2}

\vspace{0.4cm}

\begin{center}

\begin{tabular}{|c|c| c|c|c|}
\hline
\raisebox{0pt}[13pt][7pt] Experiment & Analysis & $\Delta m^2$ is ... & $\Delta m^2 ({\rm eV})^2$ & 
reference \\
\hline
\raisebox{0pt}[13pt][7pt] Kamiokande & $R$ & best fit & $1.6 \times 10^{-2}$ & 18 \\ \hline
\raisebox{0pt}[13pt][7pt] Kamiokande & up-going $\mu$ & best fit &
$3.2 \times 10^{-2}$ & 18 \\ \hline
\raisebox{0pt}[13pt][7pt] Super K & $R$ & best fit & $2.2 \times 10^{-3}$ & McGrew, Takita
\\ \hline
\raisebox{0pt}[13pt][7pt] Super K &  up-going $\mu$ & consistent with & $2.5 \times 10^{-3}$ &  McGrew, Takita
\\ \hline 
\raisebox{0pt}[13pt][7pt] Soudan II & $R$ & consistent with & $> 10^{-3}$ & Gallagher \\ \hline
\raisebox{0pt}[13pt][7pt] MACRO & up-going $\nu$ & consistent with & $5\times 10^{-3}$ & Michael
 \\ \hline 
\raisebox{0pt}[13pt][7pt] MACRO & up-going $\mu$ & consistent with & $2.5\times 10^{-3}$ & Michael
 \\ \hline 
\hline
\end{tabular}
\end{center}
\end{table*}

In 1994, the Kamiokande group reported a zenith angle dependence of
$R$.\cite{atmoskam}   Comparing the Kamioka results shown in 
Fig.~\ref{fig:k2} to the expectations in Fig.~\ref{costhetz}, the data
appear to be consistent with expectations for moderate $\Delta m^2$.
The best fit for $\nu_\mu \rightarrow \nu_\tau$ oscillations is $\Delta
m^2=1.6\times10^{-2}$ eV$^2$.

\begin{figure}
\centerline{\psfig{figure=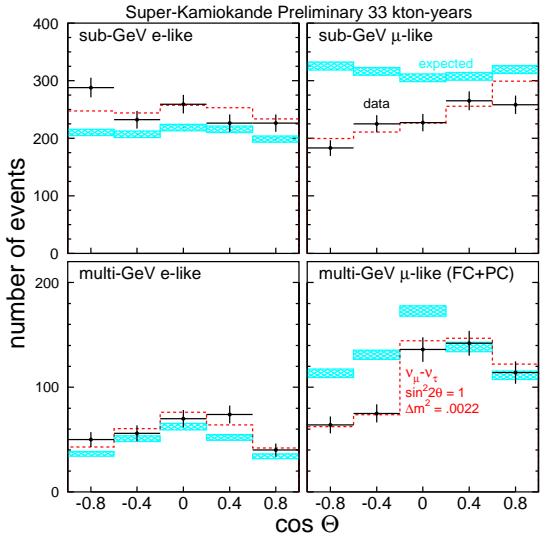,bbllx=0bp,bblly=0bp,bburx=575bp,bbury=675bp,width=3.in,clip=T}}
\caption{Rates of $\nu_e$ (``e-like'') and $\nu_\mu$ (``$\mu$-like'')
events at Super Kamiokande.  The sub-GeV sample requires $p<1.3$ GeV and
the multi-GeV sample is $p>1.3$ GeV. FC and PC refer to fully and
partially contained events.}
\label{fig:ratesk}
\end{figure}

\begin{figure}
\centerline{\psfig{figure=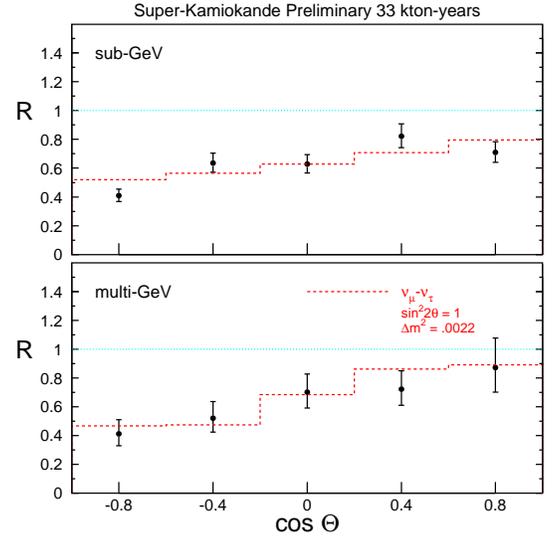,bbllx=0bp,bblly=0bp,bburx=575bp,bbury=675bp,width=3.in,clip=T}}
\caption{$R$ for the Super Kamiokande experiment.}
\label{fig:ratiosk}
\end{figure}

The zenith angle dependence has been confirmed by recent data from
Super K, as reported by C. McGrew, in these proceedings.  
The analysis is performed separately on  
sub-GeV ($p<1.3$ GeV) and multi-GeV ($p>1.3$ GeV) data samples.   The
sub-GeV are fully contained (FC) within the Super K detector while some the
multi-GeV events are only partially contained (PC).  Electron-like
($\nu_e$ scattering) candidates
and muon-like ($\nu_\mu$ scattering) candidates
are presented as a function of 
$\cos \theta_z$ in  Fig.~\ref{fig:ratesk} (points).  The shaded region 
indicates the predicted flux of neutrinos without oscillations.  There
is a further $\sim~20\%$ normalization uncertainty associated with the 
flux which is reduced to $\sim 5\%$ when the ratio, $R$, is taken. 
The dashed line, which is labled $\nu_\mu \rightarrow \nu_\tau$
for $\Delta m^2 = 0.0022 {\rm eV}^2$, shows that a good fit 
($\chi^2$/DOF=65.2/67) is achievable with an oscillation hypothesis.
Looking at the ``e-like'' data, one should note that there appears to
be a small excess in the lowest $\cos \theta_z$ bin.   Various authors
have pointed out that this allows the Super K data to accommodate 
some $\nu_\mu \rightarrow \nu_e$ at low $\Delta m^2$, as discussed in
section 4 below.   The ratio-of-ratios, $R$, for the sub- and
multi-GeV data is shown in Fig.~\ref{fig:ratiosk}.   Comparing these
results to the expectations in Fig.~\ref{costhetz}, the data are again
consistent with oscillations at moderate $\Delta m^2$.

The Soudan Experiment has observed 
a similar zenith angle dependence (see contribution by
H. Gallager).  The Soudan detector is a 1 kton fine-grained tracking
calorimeter.   While this experiment has the drawback of 
lower statistics than Super K experiment, it has the advantage of
the capability to observe the recoil proton in the neutrino interaction.
This 
substantially improves the resolution on $\theta_z$, the angle of
the incoming neutrino.   At Super K, where only the final state lepton 
is observed, the average angle between the final state lepton
direction and the incoming neutrino direction 
is 55$^\circ$ at $p=400$
MeV/c and 20$^\circ$ at 1.5 GeV/c.\cite{superk_july}   Using the
reconstructed final state particles and the outgoing lepton, the 
resolution on $\theta_z$ for Soudan is 23$^\circ$ for the 200-400 MeV data and 
8$^\circ$ for the $>600$ MeV data.   

Neutrinos which travel through the earth may
interact in the matter under the detectors.  MACRO, Super K and
Kamiokande\cite{kamup} have measured the upward-going muons from these neutrino 
interactions.   All of these experiments see rates below that which is
expected and an angular dependence which is consistent with oscillations.
In this case, a ratio to $\nu_e$
events cannot be used to reduce sensitivity to uncertainties in the
$\nu_\mu$ flux.  However, comparisons are made to a wide range for
flux models and the ratio has been found to be low in all cases.
For example, the MACRO experiment has measured a ratio of data to
Monte Carlo of $0.74 \pm 0.036{\rm (stat)} \pm 0.046{\rm (sys)}
\pm 0.013{\rm (th)}$, which includes the systematic error on the flux
calculation.   The  $\cos \theta_z < 0$ distribution, as reported by D.
Michael at ICHEP '98, is consistent with an oscillation hypothesis 
at moderate $\Delta m^2$.    As reported in the same talk, MACRO can
also measure upward-going $\nu_\mu$ interactions within the detector;
a deficit in this data, consistent with oscillations, is also observed.

\begin{figure}
\centerline{\psfig{figure=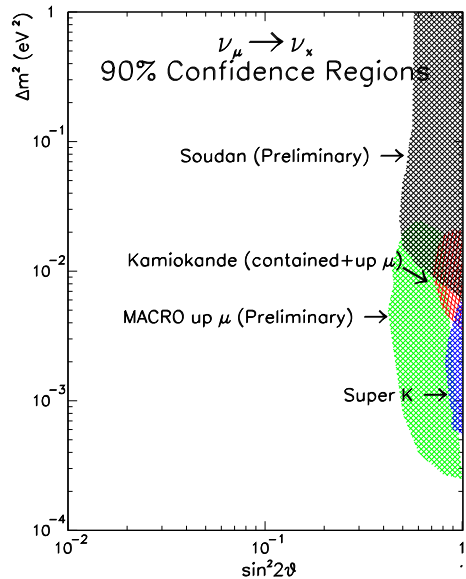,bbllx=0bp,bblly=100bp,bburx=575bp,bbury=700bp,width=3.25in,clip=T}}
\caption{A summary of 90\% CL allowed regions for Soudan II, MACRO,
Kamiokande and Super K}
\label{fig:superk_plus}
\end{figure}

Fig.~\ref{fig:superk_plus} summarizes the analyses of these
experiments under the oscillation hypothesis.  Although much attention
has been focussed on the fact that the $\Delta m^2$ range of Super K 
extends into the range of $10^{-4} {\rm eV}^2$,
it should be noted that 1) the Super K best fit is consistent
with the other results and 2) the Super K $R$ analysis integrated 
over zenith angle dependence overlaps with the $R~vs.~\cos \theta_z$
analysis only at the top of the $\Delta m^2$ range.  
Tab.~\ref{tab:deltam2}  summarizes $\Delta m^2$  for the
hypothesis that the atmospheric neutrino deficit is entirely explained
by $\nu_\mu \rightarrow \nu_\tau$ oscillations.  All results are
consistent with this oscillation hypothesis.   Experiments quote either 
a ``best fit'' or a ``consistent'' $\Delta m^2$  value, as noted.
As will be noted again in the section on Theoretical Interpretation, 
the best fits with no $\sin^2 2\theta <1$ constraint  for
Super K, Kamiokande and MACRO are all marginally in the
unphysical regions, with $\sin^2 2\theta$ of 
1.35, 1.05 and $>$1.0, respectively.  
Taken as a whole, the results are
consistent with oscillations $\Delta m^2 \sim 
2.5\times 10^{-3} {\rm eV}^2$ and $\sin^2 2\theta\sim 1$.
For a global analysis of the data from the atmospheric neutrino
experiments, see the contribution from Gonzalez-Garcia in these proceedings.

\subsection{The LSND Signal}

The LSND hint for neutrino oscillations is the only indication for
oscillations which is
a {\it signal}, as opposed to a deficit.  Evidence has been seen for
both $\bar \nu_{\mu} \rightarrow \bar \nu_e$ and $ \nu_{\mu} \rightarrow \nu_e$
oscillations.    
In 1995 the LSND experiment published data showing candidate
events that were consistent with $\bar \nu_{\mu} \rightarrow \bar \nu_e$
oscillations.\cite{paper1} Additional event excesses 
were published in 1996 and 1998 for both
$\bar \nu_{\mu} \rightarrow \bar \nu_e$
oscillations \cite{bigpaper2} and $\nu_{\mu} \rightarrow 
\nu_e$
oscillations. \cite{paper3} The two oscillation searches are
complementary, having 
different backgrounds and systematics, yet yielding 
consistent results.  

The experiment is described in the parallel
session contribution to these proceedings by R. Imlay.
This is the only oscillation signature observed from an accelerator
experiment, in this case a beam produced at LANCE at LANL, with 800
MeV energy protons interacting with 
a water target, a close-packed high-Z target and 
a water-cooled Cu beam dump. 

\begin{figure}
\centerline{\psfig{figure=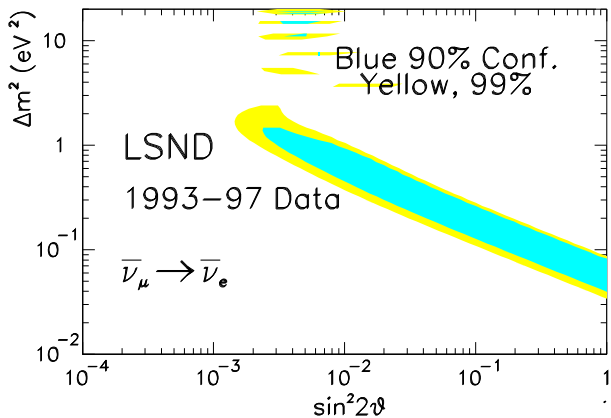,bbllx=100bp,bblly=400bp,bburx=500bp,bbury=700bp,width=3.in,clip=T}}
\caption{LSND allowed region for data from 1993-1997.} 
\label{fig:lsnd1}
\end{figure}

\begin{table*}[t]

\caption{Preliminary numbers of excess events and
the corresponding oscillation probabilities for the running periods
1993-1995, 1996-1997, and 1993-1997.}

\label{lsndtab}

\vspace{0.4cm}

\begin{center}

\begin{tabular}{|c|c|c|c|c|}
\hline
\raisebox{0pt}[13pt][7pt] Data Sample&Fitted Excess&$\bar \nu_e$ Background&Total Excess&Oscillation Probability \\
\hline
\raisebox{0pt}[13pt][7pt] 1993-1995&$63.5\pm 20.0$&$12.5\pm 2.9$&$51.0\pm 20.2$&$(0.31\pm 0.12\pm 0.05)\%
$ \\
\raisebox{0pt}[13pt][7pt] 1996-1997&$35.1 \pm 14.7$&$4.8\pm 1.1$&$30.3\pm 14.8$&$(0.32\pm 0.15\pm 0.05)\%
$ \\
\raisebox{0pt}[13pt][7pt] 1993-1997&$100.1\pm 23.4$&$17.3\pm 4.0$&$82.8\pm 23.7$&$(0.31\pm 0.09\pm 0.05)\%$ \\
\hline
\end{tabular}
\end{center}
\end{table*}

For the decay-at-rest (DAR) analysis ($\overline{\nu}_\mu \rightarrow \overline{\nu}_e$), the beam
is produced by $\pi^+$'s which stop and decay in the beam dump, producing
muons which then decay to produce $\overline{\nu}_\mu$'s. 
The signature for oscillations is a 
$\bar \nu_e$ interaction ($\bar \nu_e p \rightarrow e^+ n$), yielding
a positron signal, followed by $n p \rightarrow
d \gamma$, where the 2.2 MeV $\gamma$ is detected.  
A comparison of the  
energy dependence of the the observed $\overline{\nu}_e$ events
and the non-beam background indicates a ``Fitted Excess'' of events
as shown in Tab.~\ref{lsndtab}.
Comparing this ``Fitted Excess'' 
to the expected $\overline{\nu}_e$ beam background then 
yields a ``Total Excess'' for the 
decay-at-rest analysis, summarized in Tab.~\ref{lsndtab}, and
shown by the allowed regions in Fig.~\ref{fig:lsnd1}

\begin{figure}
\centerline{\psfig{figure=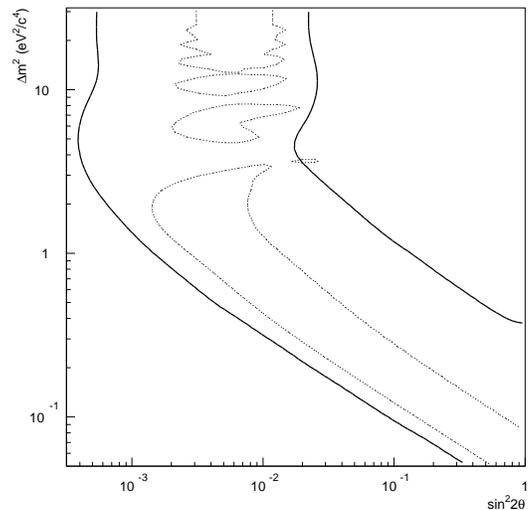,width=3.in}}
\caption{The 95\% confidence region for $\nu_\mu \rightarrow \nu_e$
oscillations (solid curve) along with the favored regions for
$\bar \nu_\mu \rightarrow \bar \nu_e$ oscillations (dotted curve).} 
\label{cl_p}
\end{figure}

The $\nu_\mu$'s in the beam from pion decay-in-flight (DIF) are also used to
probe oscillations by the LSND experiment.
The signature for $\nu_\mu \rightarrow \nu_e$ oscillations is an electron
from the reaction $\nu_e C \rightarrow e^- X$ in the energy range
$60< E_e <200$ MeV.  
The excess events are consistent with $\nu_\mu \rightarrow \nu_e$
oscillations with an oscillation probability of $(0.26 \pm 0.10 \pm 0.05)\%$.
A fit to the event distributions yields the allowed region in the
$(\sin^22\theta, \Delta m^2)$ parameter space shown in Fig. \ref{cl_p}
(between the solid lines),
which is consistent with the allowed region from the 
$\bar \nu_\mu \rightarrow \bar \nu_e$ search (dotted lines). 

In principle, when discussing signals and limits below, 
$\bar \nu_\mu \rightarrow \bar \nu_e$ should be
considered separately from  $\nu_\mu \rightarrow \nu_e$ to allow for
possible CP violation.   However, due to the good agreement between the 
the DAR ($\bar \nu_\mu \rightarrow \bar \nu_e$) and 
DIF ($\nu_\mu \rightarrow \nu_e$) data from LSND, and the lack 
of any other evidence for CP violation in neutrino oscillations, the 
two cases will be considered together below under the heading 
``$\nu_\mu \rightarrow \nu_e$.''

\section{Experiments Which Set Limits on Oscillations}

\begin{table*}[t]

\caption{ Some of the recent experiment which have set 
limits on oscillations. Regions ruled out by these experiments are 
discussed in section 3. Under ``reference'', listed names refer to
contributions to these proceedings.} 
\label{tab:exclude}

\vspace{0.4cm}

\begin{center}
{\small

\begin{tabular}{|c|c| c|c|c|c|}
\hline
Experiment & Source/Beam & $\sim E_\nu$ & $\sim L$ & 
Detector & Ref.\\
\hline \hline
CCFR/NuTeV &  accel. $\nu_\mu$, $\bar \nu_\mu$ & 100 GeV
& 1 km & Iron/scint cal and muon spect & 25, Drucker\\ \hline
Nomad & accelerator $\nu_\mu$ & 26 GeV & 1 km &
DC targ. w/i mag., EM cal, TRD & Autiero\\ \hline
CHORUS & accelerator $\nu_\mu$ &26 GeV &  1 km &
Emuls. targ. w/ scint fiber, tracking,
cal & Migliozzi \\ \hline
CDHS & accelerator $\nu_\mu$ & 1 GeV & 1 km & Iron/scint
cal and muon spect & 29 \\ \hline
BNL E776  & accelerator $\nu_\mu$ & 1.4 GeV & 1 km &
Concrete/DC cal and muon spect & 22 \\ \hline 
KARMEN 1, 2 & $\pi$ DAR $\overline{\nu}_\mu$ & 
20-60 MeV & 17 m & Liquid Scint Detector & 26, Kleinfeller\\ \hline
CHOOZ  & reactor $\overline{\nu_e}$ &  3 MeV & 1 km &
Gd-doped scintillator oil & 24 \\ \hline
Bugey & reactor $\overline{\nu_e}$ & 3 MeV & 15, 40, 95m &
Gd-doped scintillator oil & 23 \\ \hline
\end{tabular}
}
\end{center}
\end{table*}

Many experiments have searched for neutrino oscillations and seen no
signal.   Tab.~\ref{tab:exclude} provides a list of the exclusion experiments
whose results appear in this section.   Results from CCFR/NuTeV,
Nomad, CHORUS and KARMEN 2 are reviewed in these proceedings by R.
Drucker, D. Autiero, P. Migliozzi and J. Kleinfeller, respectively.

\subsection{Limits on $\nu_\mu \leftrightarrow \nu_e$ oscillations}

\begin{figure}
\centerline{\psfig{figure=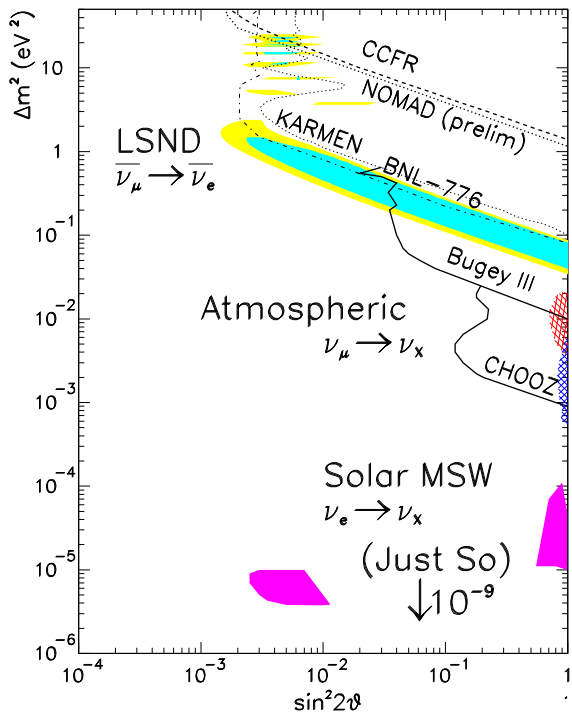,bbllx=0bp,bblly=0bp,bburx=500bp,bbury=600bp,width=3.5in,clip=T}}
\caption{Allowed and excluded regions for $\nu_\mu \leftrightarrow \nu_e$
and $\bar \nu_\mu \leftrightarrow \bar \nu_e$ oscillations.} 
\label{beforenu98}
\end{figure}

In principle, all three indications of oscillations can be
interpreted as $\nu_\mu \leftrightarrow \nu_e$.   
If the solar deficit is due to oscillations, it is entirely 
$\nu_e \rightarrow \nu_X$, where $\nu_X$ may be $\nu_\mu$ (or
$\nu_\tau$, as discussed later).   
Because of the very low $\Delta m^2$ associated with all models for
solar neutrino oscillations, no terrestrial neutrino experiment has
been able to directly test the $\nu_e \rightarrow \nu_\mu$ hypothesis.
The atmospheric deficit appears to be largely $\nu_\mu \rightarrow \nu_X$, where
$\nu_X$ could be $\nu_e$.   However, this is regarded as unlikely for
two reasons.   First, the Super K fits to the combination of
``e-like'' ($\nu_e$ candidate) and ``$\mu$-like'' ($\nu_\mu$
candidate) events give a poor $\chi^2$/DOF of 87.8/67 for 
describing the data as entirely  $\nu_\mu \leftrightarrow \nu_e$
(see contribution by McGrew, these proceedings).   Furthermore,
the reactor experiments, Bugey\cite{bugey}
and CHOOZ,\cite{chooz} which search for $\nu_e$ disappearance, have excluded
nearly all of the allowed region for atmospheric $\nu_\mu
\leftrightarrow \nu_e$.   The 90\% exclusion regions from these
experiments are indicated on Fig.~\ref{beforenu98}
LSND is required to
be $\bar \nu_\mu \rightarrow \bar \nu_e$, because it is an observed signal.   
This signal is the most amenable 
to systematic
study at accelerators due to the larger $\Delta m^2$ values involved:
$> 0.1$ eV$^2$ compared to $10^{-2} - 10^{-3}$ eV$^2$ for the
atmospheric neutrino problem and $10^{-4} - 10^{-5}$ eV$^2$ for the
solar neutrino problem. Thus in studying the LSND signal, detectors
can be placed much closer to the neutrino source and higher energy
neutrino beams can be used.  The high $\Delta m^2$ region has been
excluded by the Nomad (preliminary) and CCFR results.\cite{alex}
Therefore, interest is now focussed on the 
$\Delta m^2 = 0.1$ to $1.0~{\rm eV}^2$ region.

\begin{figure}
\centerline{\psfig{figure=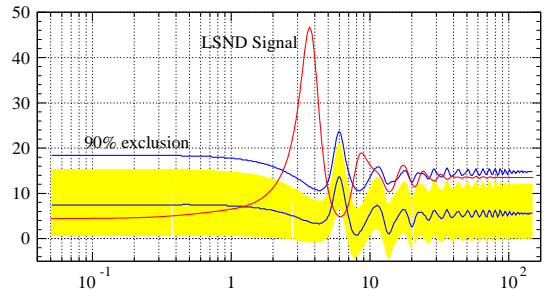,width=3.5in,clip=T}}
\caption{The KARMEN 1 excess is the central line, with one sigma
shaded errors bars, shown as a function of $\Delta m^2$. The LSND
expectation is superimposed.} 
\label{karI}
\end{figure}

The KARMEN experiment\cite{KARMEN} 
is complementary to LSND, running
at the ISIS facility in Rutherford, England, with a neutrino source produced by
pion decay-at-rest.   
The detector is smaller than LSND and the beam is less intense than at 
LANCE, resulting in lower statistics than LSND by $\sim \times 10$.  
The KARMEN 1 experiment took data through 1995.

KARMEN 1 observed an excess of approximately seven events, which is in
agreement with predictions based on the LSND central values at low
$\Delta m^2$.\cite{KARMEN} The measured KARMEN 1 excess, as a function
of $\Delta m^2$, is shown in Fig.~\ref{karI}. The shaded band showing
the error associated with the measurement indicates that the excess is
only at the $1\sigma$ level. However, the overall excess is in
agreement with the central allowed value of LSND at low $\Delta m^2$.
The LSND prediction is indicated by the labeled line in
Fig.~\ref{karI}. In the region of $\Delta m^2 < 1~{\rm eV}^2$, the
LSND central prediction of 4 events is in good agreement with the
observed excess of 7 events. Because of the low significance of the
excess, KARMEN collaboration prefers to quote a limit for their
first-run result, with the number of events set by the ``90\%
exclusion'' line shown in Fig.~\ref{karI}. The KARMEN 1 limit appears
in Fig. \ref{beforenu98} .

As reported at this conference (see contribution by J. Kleinfeller),
the second run of the KARMEN experiment (KARMEN 2) is in progress.  
Due to upgraded shielding, 
the cosmic ray background has been significantly reduced.
At this
point, $\sim 40\%$ of the data have been collected.  The background expected 
for this running period is 3 events, and a typical signal for the
$\Delta m^2 < 1 ~{\rm eV}^2$ region is 1 event, if LSND is correct.
The total number of events observed during this period (combined
signal and background) is zero (which has a 5.6\% probability of
occurring as a statistical fluctuation).  

The quandary associated with
the treatment of data in this situation has already been discussed
above in
section 1.   The limit which can be set in the situation
where an experiment fails to see the expected background will be
better than the actual experimental 
sensitivity, that is, the expectation if the
background had been observed.
Fig.~\ref{giuntips} shows various interpretations of the KARMEN 2 
null result. The Bayesian approach, the Feldman-Cousins
method and the Giunti Method give differing limits, thus making the result
hard to interpret.   The sensitivity of KARMEN 2 is shown by the lines
connected with open symbols.  The sensitivity is worse than BNL 776
and does not cover the LSND signal region.  

At this point, the situation for $\nu_\mu \leftrightarrow \nu_e$
oscillations in the LSND region can be summarized in the following manner.
Three 
results show excesses:
\begin{itemize}
\item 33.9 $\pm$ 8.0 events (LSND Decay-at-rest, $\overline{\nu}_\mu
\rightarrow \overline{\nu}_e$)
\item 18.1 $\pm$ 6.6 (LSND Decay-in-flight,  $\nu_\mu\rightarrow \nu_e$)
\item 7.3 $\pm$ 7.0 (KARMEN 1 Decay-at-rest,  $\nu_\mu\rightarrow
\nu_e$)
\end{itemize}
One experiment (KARMEN 2) which expected 3 events background, plus one event
signal, has observed zero total events.

\subsection{Limits on $\nu_\mu \leftrightarrow \nu_\tau$ oscillations}

\begin{figure}
\centerline{\psfig{figure=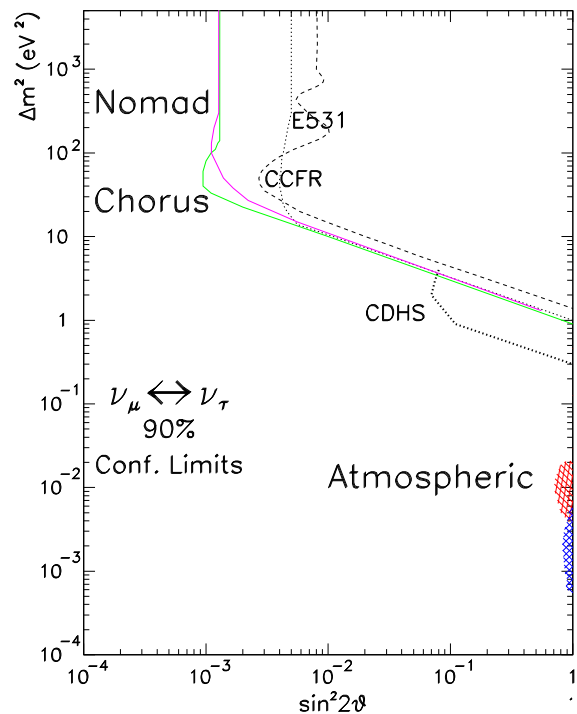,bbllx=0bp,bblly=0bp,bburx=500bp,bbury=600bp,width=3.5in,clip=T}}
\caption{Status of $\nu_\mu \rightarrow \nu_\tau$ oscillation searches.}
\label{numunutau}
\end{figure}

The CHORUS and NOMAD experiments\cite{chornom} explore the  
high $\Delta m^2$ $\nu_\mu \leftrightarrow \nu_\tau$ 
oscillation region,
where one may expect a signature from 
neutrinos which contribute to
``dark matter'' in the universe. 
Recent astrophysical data has indicated that
some of the dark matter may be ``hot.'' 
One candidate for hot dark matter is 
massive neutrinos.   
Present data can accomodate 
$\Omega = 0.1 \sim 0.4$ with $\Omega_{HDM} < 0.2$, as reviewed by B.
Kayser at this conference.
Massive neutrinos may be an important 
component of the dark matter in the
universe since the density of relic neutrinos from the Big Bang is 
$\sim$100 $\nu $'s$/$cm$^3/$type.     
Neutrinos in the mass range of $1\sim 6$ eV could help
to explain the small scale structure in the universe and recent anisotropy
measurements of the photon background radiation\cite{anisotropy}.
If one assumes that the heaviest neutrino is much more massive than
the rest, then the astrophysical data indicate 
that the region of interest for searches 
is approximately $1 < \Delta m^2 < 36~{\rm eV}^2$.

NOMAD and CHORUS share a
high-intensity $\nu_\mu $ beam produced at CERN.  The neutrino
energies range
from 10 $\sim$ 40 GeV.  The $\bar \nu_\mu$ contamination in this beam
is only $\sim 5\%$.  The prompt $\nu_\tau$ contamination is $(3 \pm
4)\times 10^{-6}$. 

The CHORUS experiment uses an 800 kg emulsion target which provides $<
1\mu$m spatial resolution.  Thus this experiment can
identify $\nu _\tau $ charged current 
interactions by seeing the $\tau$ decay in
the emulsion after a few tenths of a millimeter, producing a kink in
the track. Automatic emulsion scanning
systems have been developed to handle the large quantities of data 
($\sim 300,000$ events).  The emulsion target is followed by a magnetic
spectrometer, calorimeter and muon spectrometer allowing momentum
reconstruction and particle identification in each event.

NOMAD is a fine-grained electronic detector composed of
a large aperture dipole magnet ($3m\times 3m\times 7m$ with $B=0.4$ T)
filled with drift chambers that act as both the target and tracking medium.
The experiment uses kinematic cuts associated with the missing energy from
outgoing $\nu $'s in the $\tau $ decay to separate statistically a possible
oscillation signal. 

CHORUS and NOMAD experiments have sensitivity to oscillations
with $\Delta m^2 > 1~{\rm eV}^2$.
Preliminary negative search results are 
reported for approximately $\sin ^22\theta >2\times 10^{-3}$, as shown
in Fig.~\ref{numunutau}.     
Because no signal has been
observed in these experiments, the 
possibility that massive neutrinos are dark matter is becoming more
unlikely.  However, it remains possible that the mixing is 
very small; thus the
possibility is not entirely ruled out.

Among the hints for oscillations discussed previously, only 
the atmospheric neutrino deficit 
may result from $\nu_\mu \leftrightarrow \nu_\tau$ oscillations.
The $\Delta m^2$ reach of the present experiments does not cover 
the atmospheric allowed region.  The lowest limit, 
$\Delta m^2 \sim 0.3~{\rm eV}^2$ is
from the CDHS experiment.\cite{CDHS}

\subsection{Limits on $\nu_e \leftrightarrow \nu_\tau$ oscillations}

\begin{figure}
\centerline{\psfig{figure=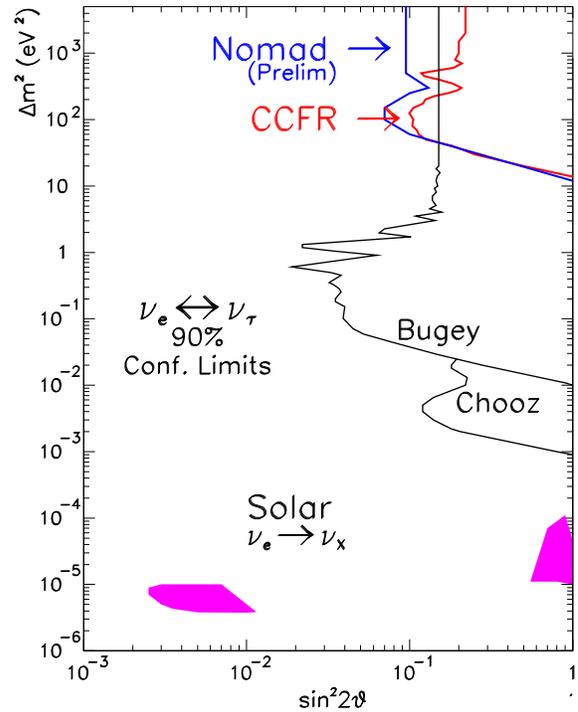,bbllx=0bp,bblly=0bp,bburx=500bp,bbury=500bp,width=3.5in,clip=T}}
\caption{Status of $\nu_e \rightarrow \nu_\tau$ oscillation searches.}
\label{nuenutau}
\end{figure}

Naively, $\nu_e \rightarrow \nu_\tau$ oscillations appear least likely,
because this skips the second generation.   However, 
this oscillation does appear in some models, as
discussed in section 4.  The one hint 
which can be interpreted as
such an oscillation is the solar neutrino deficit.   
In addition,
one or more of the neutrinos may represent 
a fraction of the dark matter in the universe.   If there is 
a mass difference between the neutrinos, this 
might manifest itself through $\nu_e \rightarrow \nu_\tau$
oscillations at relatively high $\Delta m^2$.

The excluded regions for  $\nu_e \rightarrow \nu_\tau$ are shown in
Fig.~\ref{nuenutau}, along with the solar allowed region.
Recent searches from NOMAD and CCFR have addressed high $\Delta m^2$'s,
while the reactor experiments have explored down to $10^{-3} {\rm
eV}^2$.   The terrestrial experiments remain a few orders of magnitude away
from addressing the solar $\nu_e \rightarrow \nu_\tau$ hypothesis.

\section{Theoretical Interpretation of the Data}

When comparing the evidence for oscillations with the excluded regions, 
we are faced with theoretical problems with both the suggested $\Delta m^2$
regions and the mixing angles.   There are
apparently three distinct $\Delta m^2$ regions:
\begin{eqnarray*}
\Delta m^2_{solar}&=&10^{-5}~{\rm or }~10^{-10}~{\rm eV}^2 \\
\Delta m^2_{atmos}&=&(10^{-2}~{\rm to }~10^{-4})~{\rm eV}^2 \\
\Delta m^2_{LSND} &=&(0.2~{\rm to }~2)~{\rm eV}^2 \\
\end{eqnarray*}
However, in a straightforward three-generation mixing model, if
$\Delta m^2_{12}$ is very small, then $\Delta m^2_{13} \approx  \Delta
m^2_{23}$, leading to only two apparent $\Delta m^2$ regions.
Furthermore, one must address the very large mixing in the neutrino
sector compared to the quark sector.  For the atmospheric
data there is no solution at 90\% CL with $\sin^2 2\theta$ less than 
approximately 0.6.    

The problem has been attacked in several ways.  
Several possibilities are presented in these proceedings by B. Kayser,
M. Gonzalez-Garcia, S. Nandi, and P. Roy.
Some phenomenologists 
choose to admit
complications in the simplistic three generation mixing model.
Others prefer to introduce a sterile neutrino.   Some like
to throw out the data which they like the least.   The first two
options merit further explanation.   

\subsection{Three Generation Mixing Models}

If three generation mixing is to explain all of  the data, then one of the
following must be true:
\begin{itemize}
\item $\Delta m^2_{solar} \approx \Delta m^2_{atmos}$ 
\item $\Delta m^2_{LSND} \approx \Delta m^2_{atmos}$ 
\item $\Delta m^2_{atmos}$ is a convolution of $\Delta m^2_{solar}$
and $\Delta m^2_{LSND}$.  
\end{itemize}
The first two possibilities are difficult to accommodate.  As shown in
Fig.~\ref{allhints}, the three allowed regions are well separated,
although one should keep in mind that these plots show only 1.3$\sigma$
regions.  The third possibility is now under exploration and will be
considered here.  

With three generations of neutrinos, one would expect a more complicated
oscillation phenomenology that includes transitions between all pairs.
In this case, an experiment which observes $\nu_\mu$ disappearance,
like Super K, may be seeing a combination of $\nu_\mu \rightarrow
\nu_\tau$ and $\nu_\mu \rightarrow \nu_e$.  If one then takes the data
and analyzes it in terms of only one scenario, for example only 
$\nu_\mu \rightarrow \nu_\tau$, then one will extract a $\Delta m^2$
which is some convolution of the two real $\Delta m^2$ regions.  
It is interesting to note that this could also lead to a best fit for
$\sin^2 2\theta$ which is greater than one, as has been observed by the
atmospheric experiments.

For this to be an acceptable hypothesis, the Super K data must accommodate $\nu_\mu
\rightarrow \nu_e$ oscillations at low $\Delta m^2$.  As shown in
Fig.~\ref{costhetz},  low $\Delta m^2$ oscillations will only be
apparent in the most negative $\cos \theta_z$ bins.   The Super K
data, shown in Fig.~\ref{fig:ratesk} may see an excess in
the lowest  $\cos \theta_z$ bin of the ``e-like'' data.  Therefore,
it may be possible to fit the Super-K data with an admixture of 
of $\nu_\mu \rightarrow
\nu_\tau$ and $\nu_\mu \rightarrow \nu_e$.

\begin{figure}
\centerline{\psfig{figure=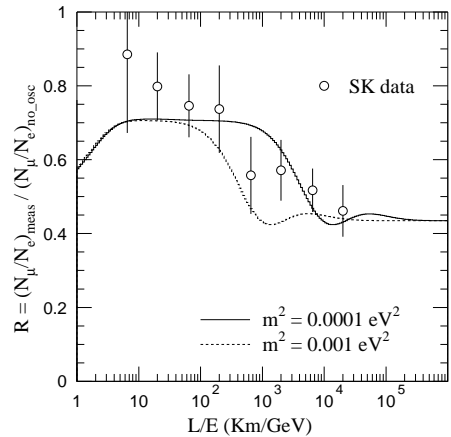,bbllx=0bp,bblly=300bp,bburx=575bp,bbury=600bp,width=3.5in,clip=T}}
\caption{Fits to the Super K atmospheric ratio-of-ratios, $R$,
assuming contributions from $\nu_\mu \rightarrow \nu_\tau$ at $\Delta
m^2 = 0.4~{\rm eV}^2$ and $\nu_\mu \rightarrow \nu_e$ at low $\Delta
m^2$.}
\label{fig:thun}
\end{figure}

As an example,
Thun and McKee have presented~\cite{ThunMcKee}    
an explanation for the 
data assuming  three generation mixing with $\Delta
m^2_1 = 0.4~{\rm eV}^2$ and  $0.0001 <\Delta m^2_2<0.001 {\rm
eV}^2$.
The mixing matrix
describing the oscillation (see equation 2) is 
\[
\left( 
\begin{array}{l}
\nu _e \\ 
\nu _\mu \\ 
\nu _\tau
\end{array}
\right) =\left( 
\begin{array}{lll}
0.78 & 0.60 & 0.18 \\ 
-0.61 & 0.66 & 0.44 \\ 
0.15 & -0.45 & 0.88
\end{array}
\right) \left( 
\begin{array}{l}
\nu _1 \\ 
\nu _2 \\ 
\nu _3
\end{array}
\right) 
\]
This choice of parameters makes certain predictions for
each of the three indications of neutrino oscillations:
\begin{description}
\item[atmospheric data --]  Using $\Delta m^2_2 = 3 \times 10^{-4}{\rm eV}^2$ for an
example, this model predicts the ratio-of-ratios for the atmospheric
data to be 0.72, which is consistent with the experimental average of 
$\sim 0.6\pm0.1$.  As shown in Fig.~\ref{fig:thun}, the model fits the 
$\cos \theta_z$ dependence of the Super K $R$ reasonably well for the  
quoted range of $\Delta m^2_2$, although the normalization appears to
be low.  This fit would improve if the 
higher $\Delta m^2$ value were chosen to be 0.2 eV$^2$, leading to a
shift upward in the overall normalization.  
\item[solar data --] The Thun-McKee solution 
predicts that the ratio of data to Monte Carlo in the solar
experiments should be 0.5 for all experiments.   For this to be
correct, then the Homestake experiment would have to be observing a
ratio which has fluctuated low by 3.1$\sigma$ if the BP98 (SSM) model is used.  
However, if the ratio is taken using the 
Turck-Chieze and Lopes solar model\cite{Turck}, then one obtains $0.403\pm0.025\pm0.025$
which is in agreement with the Thun-McKee assumption of 0.5.   The
choice of large-mixing angle solar solution is somewhat disfavored
unless one introduces an increase in the {\it hep} neutrinos, as
discussed in section 2, above.    
\item[LSND data --] The choice of 
$\Delta m^2 = 0.4~{\rm eV}^2$ and $\sin^2 2\theta_{eff} = 0.026$
for LSND is within the allowed region, although extremely close to the
Bugey 90\% CL limit.
These parameters indicate that the KARMEN 1
experiment would see 3 event excess, which is consistent with the 
measured 7 event excess.   For the total KARMEN 2 running, this model
predicts a 2 event excess, hence to date less than 1 event is expected.
At this point, KARMEN 2 has failed to observe either an excess or the
expected three background events (see section 3).  
\end{description}
There is no quoted overall $\chi^2$ for the Thun and Mckee (nor any
other) three-generation model. 

It is important to recognize that Thun and McKee have not fit the
data. They have simply chosen a set of parameters and demonstrated
that it is possible to develop a model which fits much of
the experimental data.  Other models of three generation mixing 
have also been developed based on a clever choice of parameters
which also come close to fitting all of the data.\cite{teshima}
In every case, there are experimental issues
which make the choice of parameters uncomfortable.   A global analysis, which
incorporates all of the data including systematic errors and
correlations, similar to the fits done to obtain the electroweak
parameters, is needed.

\subsection{Sterile Neutrinos}

The second option for fitting the experimental data is to introduce a
sterile neutrino.  The sterile neutrino $\nu_s$ does not interact
through the weak interaction because it is postulated to be an isosinglet partner to the 
``standard light neutrinos.'' Thus oscillations between a
$\nu_e$, $\nu_\mu$ or $\nu_\tau$  and a $\nu_s$ would cause the
standard neutrino to apparently disappear.  Obviously sterile
neutrinos can only be invoked for cases where a deficit (as opposed to
a signal) are observed.  So sterile neutrinos may provide the
explanation for the atmospheric or solar deficits, but LSND is
required to be  $\nu_\mu \rightarrow \nu_e$.

Sterile neutrinos 
were first proposed as an explanation for the solar deficit 
by Caldwell and Mohapatra\cite{Caldwell} in 1993.  A more recent
example is the model by Barger,
Weiler and Whisnant.\cite{Barger}  In this picture, the solar deficit is
explained by $\nu_e \rightarrow \nu_s$ oscillations.   The atmospheric
deficit is  $\nu_\mu \rightarrow \nu_\tau$.   

Alternatively, the
atmospheric result may be largely $\nu_\mu \rightarrow\nu_s$,
as suggested by Joshipura and Smirnov\cite{Smirnov}.   The solar
data can then be explained as a combination of $\nu_e \rightarrow
\nu_\tau$ and $\nu_e \rightarrow \nu_s$.  

The sterile neutrino has several nice features.  By adding an extra
degree of freedom (the mass of the $\nu_s$) to the theory, one can 
comfortably fit all of the data.   The $\nu_s$ provides an isosinglet
for grand unified theories, however the apparent light mass of the
$\nu_s$ presents difficulties.  The extremely large apparent mixing
angles can be explained by arguing that the $\nu_e \leftrightarrow
\nu_\mu \leftrightarrow \nu_\tau$ mixings are comparable to the quark
sector while mixing to $\nu_s$ is large.   Finally, the $\nu_s$ is a
candidate for hot dark matter. 

\section{Future Experiments}

The existing indications of neutrino oscillations raise many questions for
future experiments to address.   Many new experiments are proposed 
to address the issues which have been raised by the present data.
This section provides an overview of some of the exciting results
which can be expected in the near future.

\subsection{Future Tests of Solar Neutrino Oscillations}

\begin{table*}

\caption{Some future Solar Neutrino Experiments}
\label{futuresun}

\begin{center}

\begin{tabular}{|c|c|c|c|c|}
\hline
Experiment & Detector & Search & Source & Approx. Range \\ \hline
\hline
SNO & Deuterium & $\nu_x + d \rightarrow p + n + \nu_x$ (NC $x=e,\mu,\tau$)
types) & Sun & $>5 MeV$ \\
    &           & $\nu_e + d \rightarrow p + p + e^-$ (CC $\nu_e$
only) &  &  \\ \hline
BOREXINO & Liquid Scint. & $\nu_e$ elastic scatters & Sun & $0.5<E<1.0$ MeV \\ \hline
GNO & Gallium & $\nu_e$ capture in $Ga$ & Sun & $>0.2$ MeV \\ \hline
HELLAZ & Helium TPC & $\nu_e$ elastic scatters & Sun & $> 0.05$ MeV
\\ \hline \hline
KamLAND  & Liquid Scint. & $\nu_e$ elastic scatters &
 Reactors & $>1$ MeV \\ \hline
\end{tabular}

\end{center}

\end{table*}

The issues related to the solar neutrino deficit are:
\begin{itemize}
\item Can we see the $L/E$ dependence which will clearly demonstrate 
neutrino oscillations?
\item Is this $\nu_e \rightarrow \nu_\mu$, $\nu_\tau$ or $\nu_s$ (or
some combination)?
\item What is the $\Delta m^2$?  Is MSW or Just-So the right
explanation?
\item If the solution is MSW, is it the small or large angle solution?
\item Is there any room for doubting the Standard Solar Model?
\end{itemize}

In order to address the $L/E$ dependence, a wide range of experiments 
with varying energy thresholds is needed.  Tab.~\ref{futuresun}
provides a summary of the upcoming solar
experiments which were presented at ICHEP'98 (see contributions by C. Waltham,
M. Chen, T. Patzak, D. Vignaud 
and Y.F. Wang, these proceedings).  As can be seen from the Tab.~\ref{futuresun}
these and other proposed future experiments will cover energies
ranging upward from 0.05 MeV, permitting tests of the $L/E$ dependence.

The BOREXINO experiment\cite{Borexino}
is sensitive to neutrinos from the $^7Be + e^{-} \rightarrow ^7{Li} + \nu_e$ 
interaction, which is a delta function in the flux distribution,
as shown on Fig.~\ref{nuesun}.  Therefore, this experiment will be
highly sensitive to seasonal variations in $L$, the earth-sun
distance, 
if the ``Just So'' 
solution is correct.

The SNO experiment\cite{SNO} will test the hypothesis for 
sterile neutrino solar oscillations.
This experiment observes neutral current (NC) interactions for
all three standard neutrinos and charged
current (CC) $\nu_e$ interactions.  Sterile neutrinos will not have neutral
current interactions.   Therefore the ratio of NC to CC interaction
rates will be lower than predicted if solar oscillations are $\nu_e
\rightarrow \nu_s$.

As the only terrestrial experiment which can
address the solar neutrino question, the KamLAND experiment\cite{Kamland} 
is not affected by theoretical errors from the Standard Solar Models. 
This experiment will be located in the Kamiokande cavern and will make use of
neutrinos from five reactor sites, resulting in $L \sim $ 160 km.
This experiment is sensitive only to the LMA MSW solution.

In the far future, an interesting test of LMA MSW $\nu_e \rightarrow
\nu_\tau$ and $\nu_e \rightarrow \nu_\mu$
oscillations may be made by the first stage of the muon collider.  A
muon storage ring would provide an intense beam of $\nu_e$'s (and
$\nu_\mu$'s) from muon decays.   Using a 50 GeV storage ring and beams
directed from the US to Italy and Japan, single event sensitivity
covers the upper region of the LMA solution.\cite{mucollide}

\subsection{Future Tests of Atmospheric Neutrino Oscillations}

The issues related to the atmospheric neutrino deficit are:
\begin{itemize}
\item Can we see an effect in the controlled environment of an
accelerator experiment?
\item Is this mainly $\nu_\mu \rightarrow \nu_\tau$ or $\nu_s$ (or
some combination)?
\item Is there any $\nu_\mu \rightarrow \nu_e$ component?
\item What is the $\Delta m^2$?
\end{itemize}

\begin{figure}
\centerline{\psfig{figure=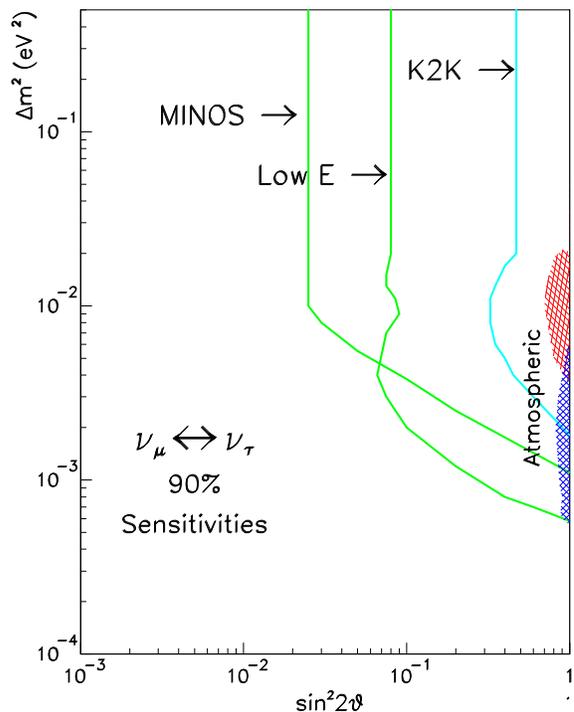,bbllx=0bp,bblly=0bp,bburx=500bp,bbury=600bp,width=3.5in,clip=T}}
\caption{Sensitivites for MINOS and K2K, two long baseline experiments
approved to run in the near future.  MINOS is also considering running with 
a low energy (``Low E'') beam design.}
\label{fig:longbase}
\end{figure}

It is possible to be sensitive to 
the moderate $\Delta m^2$'s indicated by the atmospheric neutrino
deficit through ``long baseline'' experiments.   In the near future,
beams will be built at accelerator facilities with $E_\nu\sim 1-10$ GeV 
which point to detectors at  distances of hundreds of kilometers.  This
opens a new era of tests of neutrino oscillations in the atmospheric
region, with entirely different systematics from the previous
experiments.

The sensitivities of two long baseline experiments which 
are approved to run in the near future are shown in Fig.~\ref{fig:longbase}.
The K2K experiment, which uses a 250 km baseline from KEK to the
Super K detector,
will begin taking data in 1999.   The MINOS experiment, with a 730 km
baseline from FNAL to Minnesota, will begin taking data in 2002.   The
90\% CL expectations for $\nu_\mu$ disappearance are shown in Fig.~\ref{fig:longbase}.The
MINOS experiment has both a ``standard'' and ``low energy'' beam
configuration, as indicated by the two lines.  Each of these
experiments has a near detector to measure the initial beam flux.  The
far detectors are sensitive to $\nu_\mu$ disappearance and $\nu_e$
appearance.   The K2K experiment is sensitive to the full Kamiokande
allowed region.\cite{K2K}   If the three generation mixing scenario with 
$\Delta m^2 \sim 0.3~{\rm eV}^2$ $\nu_\mu \rightarrow \nu_\tau$
oscillations is correct, then K2K will see a deficit.    The MINOS
``standard beam'' configuration  covers the region where all of the
atmospheric experiments overlap at greater than 5$\sigma$.   Full
coverage of the Super K region is obtained with the ``low energy''
beam configuration.\cite{MINOSTDR}

CERN is in the process of planning an extensive long baseline program
using a beam directed to the Gran Sasso facility.  The richness of
this program lies in the diversity of detectors, addressing important
issues of systematics and allowing both appearance and disappearance
studies.   The ICARUS experiment has been approved 
and is presently under construction.   The 600 ton liquid argon
calorimeter, to be completed in 2000, will be sensitive to $\nu_\mu$
disappearance and $\nu_e$ appearance.\cite{ICARUS} The NOE experiment,
consisting of scintillating fibers alternating with TRD detectors,
also will have  $\nu_\mu$
disappearance and $\nu_e$ appearance capabilities.\cite{NOE}
The AQUARICH experiment proposes a 27-kton water target/ring-imaging
Cerenkov detector, allowing a search for $\nu_\mu$
disappearance.\cite{aquarich}   The baseline from CERN to the Gran
Sasso is 740 km.  Various beam energy options are under discussion.

\begin{figure}
\centerline{\psfig{figure=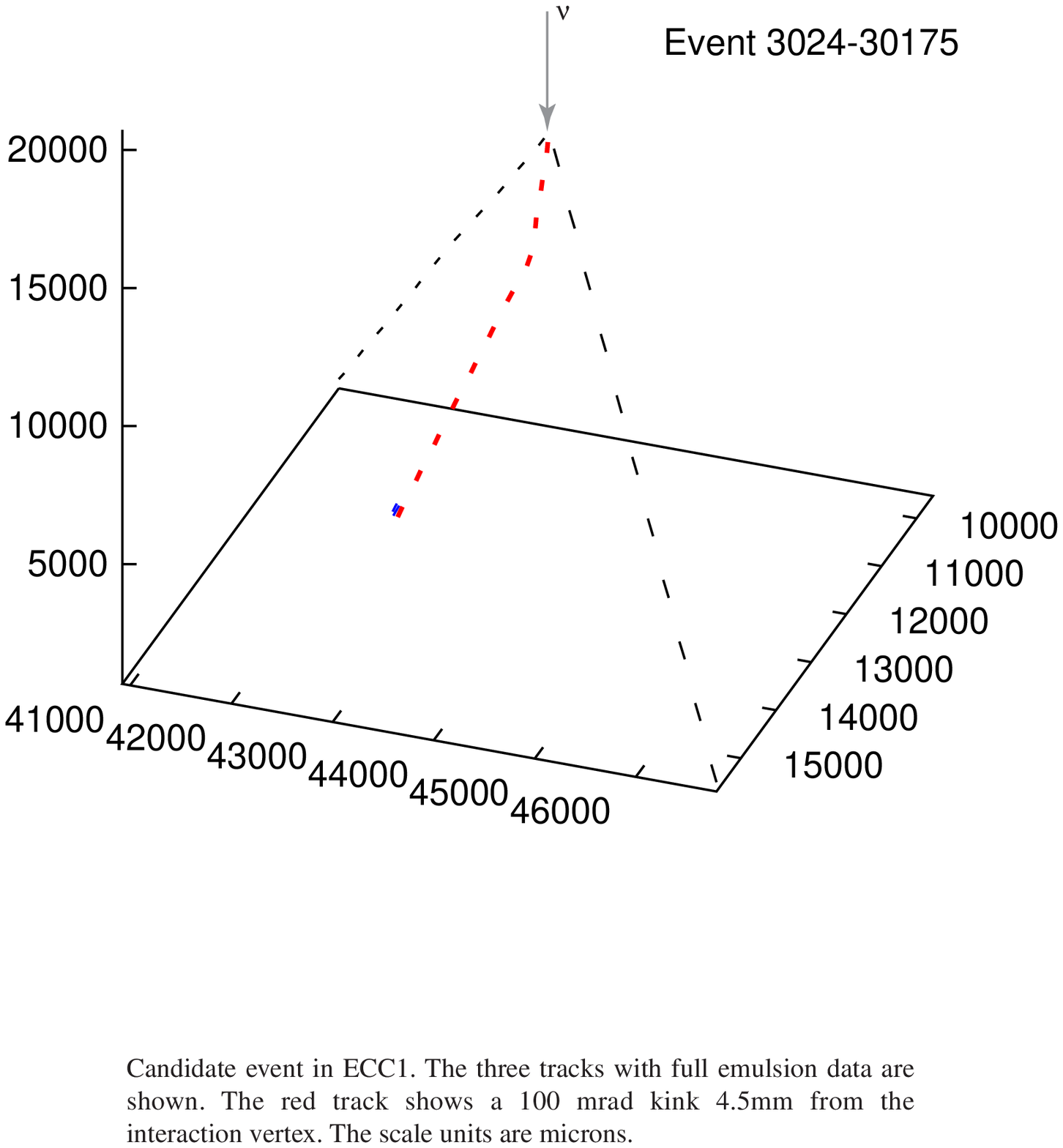,bbllx=50bp,bblly=200bp,bburx=600bp,bbury=750bp,width=3.5in,clip=T}}
\caption{$\nu_\tau$ interaction candidate observed by the DONUT
experiment at Fermilab.  The neutrino vector is indicated by the
arrow.  The interaction produced three tracks in the emulsion.  One
track shows a distinct kink, which is the signature of a $\tau$ decay.}
\label{DONUTpic}
\end{figure}

Perhaps the most exciting future possibilities lie with experiments which can
detect $\tau$'s produced in $\nu_\mu \rightarrow \nu_\tau$
oscillations.  Observation of a significant signal would be decisive.  
The most promising detector technology for $\tau$
identification is emulsion.   Using a beam dump neutrino source, 
the DONUT (``Discovery Of the NU Tau'') Experiment at FNAL recently 
identified several $\nu_\tau$ interaction candidates, thus 
possibly providing the first
direct observation of $\nu_\tau$ and also demonstrating the capability
of hybrid emulsion detectors.\cite{DONUT}  A candidate $\nu_\tau$
event is shown in Fig.~\ref{DONUTpic}.  The MINOS experiment using the
FNAL beam and the OPERA experiment using the CERN beam are considering
running with $\sim 1$ kton emulsion detectors.  Using MINOS as an
example, one would expect 7 events at $\Delta m^2 = 1 \times 10^{-3}
{\rm eV}^2$ in two years of running with the standard beam
configuration.  

The TOSCA experiment at CERN is a proposed short baseline emulsion
experiment.  This experiment is exploring $\nu_\mu \rightarrow\nu_\tau$ 
oscillations in the high 
$\Delta m^2$ region.   It was
originally motivated by the hot dark matter neutrino scenario. 
If TOSCA were to observe oscillations in the $\Delta m^2 > 1 {\rm
eV}^2$ and small mixing angle region, this 
would render $\nu_\mu \rightarrow \nu_s$ the most likely
explanation
for the atmospheric neutrino deficit.   TOSCA will have sensitivity 
down to $\Delta m^2 \sim 0.4~{\rm eV}^2$ region, also allowing this
experiment to investigate three-generation mixing models incorporating
the atmospheric result, such as the one proposed by Thun and McKee
(see section 4).

\subsection{Future Tests of the LSND Signal}

The issues related to the LSND signal are:
\begin{itemize}
\item Is the signal due to oscillations?
\item What is the $\Delta m^2$?
\item What is the $\sin^2 2\theta$?
\end{itemize}
At this point enticing signals have been seen in three types of
experiments exploring this channel (see section 2).  
What is required at this point is an
experiment which definitively covers the entire 
LSND allowed region at $>5\sigma$
and which can accurately measure the oscillation parameters.

BooNE (Booster Neutrino Experiment), which has been approved at FNAL, 
will be capable of observing both
$\nu_\mu \rightarrow \nu_e$ appearance and $\nu_\mu$ disappearance.
The first phase, MiniBooNE, is a single
detector experiment designed to
obtain $\sim 1000$ events per year if the
LSND signal is due to $\nu _\mu \rightarrow \nu _e$ oscillations.
This establishes the oscillation signal at the $\sim 8 \sigma$ level.
The second phase of the experiment introduces a second detector, with the
goal to accurately measure the $\Delta m^2$ and $\sin ^22\theta $
para\-meters of observed oscillations.

The MiniBooNE experiment\cite{mBooNE} (phase 1 of BooNE) 
will begin taking data in 2001.
The detector
will consist of a spherical tank
6 m in radius. An inner structure at 5.5
m radius will support 1220 8-inch phototubes
(10\% coverage) pointed inward and optically isolated from the
outer region of the tank.  The vessel will be filled with
769~ton of mineral oil, resulting in a 445 ton fiducial volume.
The outer volume will
serve as a veto shield for identifying particles both entering and leaving
the detector, with 292 phototubes mounted on the support structure
facing outwards.  
The first detector will be located 500 m
from the Booster
neutrino source.
The neutrino beam, constructed using the 8 GeV proton Booster at FNAL,
will have an average beam energy of approximately 0.75 GeV.

\begin{figure}
\centerline{\psfig{figure=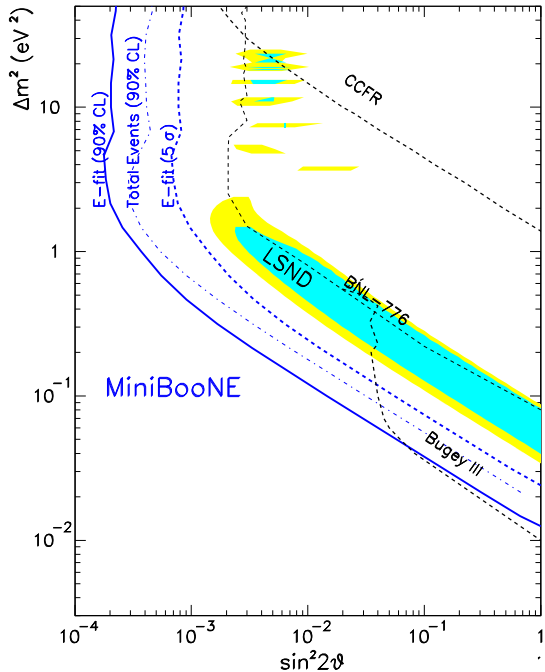,bbllx=100bp,bblly=120bp,bburx=550bp,bbury=650bp,width=3.in}}
\caption{Sensitivity regions for the MiniBooNE experiment with 
$5 \times 10^{20}$ protons
on target (1 year).  The solid (dashed) curve is the 90\% CL ($5 \sigma$)  
region using the energy fit method and 
the dashed-dot curve is the 90\% CL region using the total event method.}
\label{new_limit_plot}
\end{figure}

The sensitivity to oscillations can be calculated by summing over
energy or by including energy dependence in the fit.   
As shown in Fig. \ref{new_limit_plot}, both the summed analysis and 
the energy-dependent analysis extend well beyond the LSND allowed
region at 90\% CL.
Also shown is the
region where MiniBooNE will see a $5\sigma $ or greater signal above
background and make a conclusive measurement, which again extends
well beyond the LSND signal region.

\section{Conclusions}

The results which were presented in parallel session 2 on neutrino
oscillations are both exciting and confusing.   The three hints for
neutrino oscillations -- solar, atmospheric and LSND -- are all at the 
edge of being conclusive.   
If confirmed, this would mean
that neutrinos have mass and that lepton number is not completely
conserved -- thereby changing our understanding of the Standard Model.
Yet we cannot explain why there should be
three significantly different $\Delta m^2$ regions.  Nor do we
understand the large mixing angles.  We may be observing 
oscillations to an entirely new particle, the sterile neutrino.  
Many new measurements are planned or underway and have the goal to tell
definitively if any of the hints are indeed associated with neutrino
oscillations.
There is much left to learn before we agree on the answers to the
questions presented at ICHEP'98.  
But one thing is without question: these are exciting times for physics!

\section*{Acknowledgements}
Thanks to all of the speakers from ICHEP '98
Parallel Session 2, and to Bill Louis and Mike Shaevitz.


\section*{References}

\begin{thebibliography}{99}

\bibitem{pdg}  R. M. Barnett, {\it et.\ al.\ }, E. Phys. J. C {\bf 3}
23 (1998).

\bibitem{oldpdg} For example, R. M. Barnett, {\it et.\ al.\ }, Phys.\ Rev.\
D {\bf 54}, 162 (1996).

\bibitem{feldman}
G. J. Feldman and R. D. Cousins, 
Phys.\ Rev.\ D {\bf57} 3873 (1998).

\bibitem{giunti}
C. Giunti, hep-ph/9808240, (1998).

\bibitem{BP98} 
J.N. Bahcall, S. Basu and M. H. Pinsonneault, Phys.\ Lett.\ B {\bf
433} 1 (1998).


\bibitem{davis}
R. Davis, Prog. Part. Nucl. Phys. {\bf 32}, 13 (1994).

\bibitem{gallium}
J. N. Abdurashitov {\it et.\ al.\ }, Phys.\ Lett.\ B {\bf 328}, 234 (1994);
P. Anselmanni {\it et.\ al.\ }, Phys.\ Lett.\ B {\bf 328}, 377 (1994).

\bibitem{skamsun}
Y. Fukuda {\it et.\ al.\ }, Phys. Rev. Lett. {\bf 77} 1683 (1996).

\bibitem{Bahcall_hep} J. N. Bahcall and P. I. Krastev,
Phys. Lett. B {\bf 436} (243) 1998.


\bibitem{Adelberger} E. G. Adelberger {\it et.\ al.\ }, ``Solar Fusion
Cross Sections,'' To be published in Rev. \ Mod. \ Phys., Oct. 1998, 
astro-ph/9805121. 

\bibitem{bahcallseis} For example, J. Christensen-Dalsgaard {\it
et.\ al.\ }, Science {\bf 272} 1286 (1996).


\bibitem{helioseis}  For example, 
Castellani {\it et.\ al.\ }, Nucl. Phys. Proc. Suppl. {\bf 70} 301 (1998). 

\bibitem{hata}
N.\ Hata and P.\ Langacker,
Phys. Rev. D {\bf 56} 6107 (1997); 
N.\ Hata and P.\ Langacker,
Phys. Rev. D {\bf 50}, 632 (1994).


\bibitem{wolf} L. Wolfenstein, Phys. Rev. D17, 2369 (1978); D20, 2634 (
1979); S. P. Mikheyev and A. Yu. Smirnov, Yad. Fiz. 42, 1441 (1985) [Sov. J.
                    Nucl. Phys. 42, 913 (1986)]; Nuovo Cimento 9C, 17 (1986).


\bibitem{BKS} J.N. Bahcall, P.I. Krastev, and A. Yu. Smirnov,
hep-ph/9807216, submitted to Phys. Rev. D.


\bibitem{atmoskam}
Y.\ Hirata {\it et.\ al.\ }, Phys. Lett. B{\bf 335}, 237 (1994).

\bibitem{superk_july}
Y. Fukuda {\it et.\ al.\ }, Phys. Rev. Lett. {\bf 81} 1158 (1998).

\bibitem{kamup}
S. Hatakeyama {\it et.\ al.\ }, Phys. Rev. Lett {\bf 81} 2016 (1998). 


\bibitem{paper1}
C.\ Athanassopoulos {\it et.\ al.\ }, Phys.\ Rev.\ Lett. {\bf 75}, 2650 (1995);


\bibitem{bigpaper2}
C.\ Athanassopoulos {\it et.\ al.\ }, Phys.\ Rev.\ Lett. {\bf 77}, 3082 (1996);
C.\ Athanassopoulos  {\it et.\ al.\ }, Phys.\ Rev.\ C. {\bf 54}, 2685 (1996).



\bibitem{paper3} C.\ Athanassopoulos {\it et.\ al.\ }, LA-UR-97-1998,
submitted to Phys.\ Rev.\ C.


\bibitem{E776} L. Borodovsky {\it et.\ al.\ }, Phys. Rev. Lett. {\bf
68}, 274 (1992).

\bibitem{bugey}
B.\  Achkar {\it et.\ al.\ }, Nucl. Phys. {\bf B434}, 503 (1995).

\bibitem{chooz}
M.\  Apollonio {\it et.\ al.\ }, hep-ex/9711002, submitted to Phys. Lett. B.


\bibitem{alex} A.~Romosan {\it et.\ al.\ }, Phys. Rev. Lett. 
{\bf 78} 2912 (1997).

\bibitem{KARMEN}
B.\ Bodmann {\it et.\ al.\ }, Phys.\ Lett.\ B {\bf 267}, 321 (1991);
B.\ Bodmann {\it et.\ al.\ }, Phys.\ Lett.\ B {\bf 280}, 198 (1992);
B.\ Zeitnitz {\it et.\ al.\ },
Prog. Part. Nucl. Phys., {\bf 32} 351 (1994).
K. Eitel, hep-ex/9706023.


\bibitem{chornom}  CHORUS WWW Site:\\ 
http://choruswww.cern.ch/\\
NOMAD WWW Site: http://nomadinfo.cern.ch/


\bibitem{anisotropy}  J.R. Primack, {\it et al.}, Phys. Rev. Lett. 74 (1995)
2160.

\bibitem{CDHS} F.\ Dydak {\it et.\ al.\ }, Phys.\ Lett.\ B {\bf 134}, 281 (1984).


\bibitem{ThunMcKee} R.P. Thun and S. McKee, hep-ph/9806534, submitted
Phys. Lett. B. 

\bibitem{Turck} S. Turck-Chieze and I. Lopes, Astrophys. J. 108 (1993).

\bibitem{teshima} Two other examples are: G. Barenboim and F. Scheck,
hep-ph/9808327; T. Teshima and T. Sakai, hep-ph/9805386.

\bibitem{Caldwell} D.Caldwell and R. Mohapatra, Phys. Rev. D {\bf 48} 
3259 (1993).

\bibitem{Barger} V. Barger, T.J. Weiler, and K. Whisnant,
hep-ph/9807319 (1998).

\bibitem{Smirnov} A. S. Joshipura and A.Yu. Smirnov, hep-ph/9806376.

\bibitem{Borexino} See http://almime.mi.infn.it/

\bibitem{SNO} See http://www.sno.phy.queensu.ca/

\bibitem{Kamland}  Y. F. Wang, STANFORD-HEP-98-04 (1998). 
P. Alivisatos {\it et al.}, STANFORD-HEP-98-03 (1998).

\bibitem{mucollide} S. Geer, Phys. Rev. D {\bf 57} 6989 (1998). 
S. Geer, FERMILAB-CONF-97-417.

\bibitem{K2K} Y. Oyama, hep-ex/9803014 (1998).

\bibitem{MINOSTDR} 
See http://www.hep.anl.gov/NDK/\\
HyperText/minos$\underline{~}$tdr.html.  
P875, ``A Long Baseline Neutrino Oscillation
Experiment at Fermilab, February, 1995;
Dave Ayres for the MINOS collaboration, ``Summary of the MINOS proposal,
A Long Baseline Neutrino Oscillation Experiment at Fermilab", March
1995; NuMI note: NuMI-L-71, see: \\
http://www.hep.anl.gov/NDK/Hypertext/\\
numi$\underline{~}$notes.html


\bibitem{ICARUS} See
http://www.aquila.infn.it/icarus/.
C. Rubbia {\it et.\ al.\ }, LNGS-94/99 Vol. I \& II, (1994).
C. Rubbia, Nucl. Phys. Proc. Suppl. 48 172 (1996). 

\bibitem{NOE} See: http://www1.na.infn.it/wsubnucl/accel/noe/noe.html


\bibitem{aquarich}
T. Ypsilantis {\it et.\ al.\ }, Nucl. Instrum. Meth. {\bf A} 371 330 1996 

\bibitem{DONUT} See http://fn872.fnal.gov/ and\\
http://fn872.fnal.gov/other/files/UsersMtg98.html.

\bibitem{mBooNE} See: http://www.neutrino.lanl.gov/BooNE/\\
E. Chruch {\it et.\ al.\ } FERMILAB-P-0898 (1997).

\end{thebibliography}
\end{document}